\documentclass[11pt,a4paper]{article}

\usepackage{jheppub}
\usepackage{pdfsync}

\usepackage{epsfig,epsf}
\usepackage{amsmath}
\usepackage{amsthm}
\usepackage{amsfonts}
\usepackage{amssymb}
\usepackage{dsfont}
\usepackage{marvosym}
\usepackage{multirow}
\usepackage{mathrsfs}

\RequirePackage{graphicx}
\RequirePackage{flushend}
\RequirePackage[numbers,sort&compress]{natbib}

\usepackage{slashed}
\usepackage{epstopdf}
\usepackage[active]{srcltx}

\def\II{\hbox{{1}\kern-.25em\hbox{l}}}

\DeclareMathOperator{\Li}{Li}
\title{Next-to-leading-power kinematic  corrections to DVCS: \\  a scalar target}

\author[a]{V. M. Braun,}

\author[b]{Yao Ji,}

\author[c]{and A. N.  Manashov}

\subheader{
\begin{flushright}
{\small
TUM-HEP-1432/22 \\
DESY-22-169
}
\end{flushright}
}

\affiliation[a]{
   Institut f\"ur Theoretische Physik, Universit\"at
   Regensburg,  D-93040 Regensburg, Germany}

\affiliation[b]{Physik Department T31, James-Franck-Stra\ss e 1,
   Technische Universit{\"a}t M{\"u}nchen,
   D-85748 Garching, Germany
}

\affiliation[c]{
   II. Institut f\"ur Theoretische Physik, Universit\"at Hamburg,
   D-22761 Hamburg, Germany}

\emailAdd{vladimir.braun@physik.ur.de}

\emailAdd{yao.ji@tum.de}

\emailAdd{alexander.manashov@desy.de}

\abstract{
            Using the recent results 
            on the contributions
of descendants of the leading twist operators to the operator product expansion
of two electromagnetic currents  we derive explicit expressions
for the kinematic finite-$t$ and target mass corrections to the DVCS helicity amplitudes
to the  $1/Q^4$ power accuracy.
The cancellation of IR divergences for kinematic corrections is demonstrated to all powers in the leading order of perturbation theory.
We also argue that target mass corrections in the coherent DVCS
from nuclei are small and do not invalidate the factorization theorem.
         }


\begin{document}

\maketitle

%\newpage

%%%%%%%%%%%%%%%%%%%%%%%%%%%%%%%%%%%%%%%%%%%%%%%%%%%%%%%%%%%%%%%%%%%%%%%%%%%%%%%%%%%%%%%%%%%%%%%%%%%%%%%%%%%%%%%%%%%%%%%%%%%%%%%%

\section{Introduction}\label{sec:intro}

%%%%%%%%%%%%%%%%%%%%%%%%%%%%%%%%%%%%%%%%%%%%%%%%%%%%%%%%%%%%%%%%%%%%%%%%%%%%%%%%%%%%%%%%%%%%%%%%%%%%%%
%%%%%%%%%%%%%%%%%%%%%%%%%%

A three-dimensional ``tomographic'' imaging of the proton and light nuclei is an active research topic and a major science goal for the
planned  Electron-Ion Collider (EIC) \cite{AbdulKhalek:2021gbh,AbdulKhalek:2022erw}. Studies of the  deeply-virtual Compton scattering
(DVCS) play an important role in this undertaking. This reaction gives access to the generalized parton distributions
(GPDs)~\cite{Muller:1994ses,Ji:1996nm,Radyushkin:1997ki}  that encode the information on the transverse position of quarks and gluons in
the proton in dependence on their longitudinal momentum. This process will be measured with very high precision and in a broad kinematic
range. The QCD description of the DVCS is based on collinear factorization with GPDs as nonperturbative inputs and coefficient functions
(CFs) which can be calculated order by order in perturbation theory. At leading power, the complete next-to-leading-order (NLO) results are
available since many years~\cite{Ji:1998xh,Belitsky:1999hf,Belitsky:1998gc,Noritzsch:2003un}, and the work is ongoing to extend this
description to NNLO~\cite{Kumericki:2006xx,Kumericki:2007sa,Braun:2017cih,Braun:2020yib,Braun:2021grd,Gao:2021iqq,
Braun:2022byg,VanThurenhout:2022hgd,Braun:2022bpn}.

Beyond the leading twist, power-suppressed contributions $\sim (\sqrt{-t}/Q)^k $ and $\sim (m/Q)^k$
where $t$ is the invariant momentum transfer and $m$ is the target mass, 
have to be taken into account. 
The spatial position of partons is Fourier conjugate to the momentum transfer to the
nucleon in the scattering process. Hence the resolving power of DVCS is directly limited by the range of the invariant moment transfer $t$
available in the analysis. For the stated goal of the three-dimensional imaging, theoretical control over power corrections
$(\sqrt{-t}/Q)^k$ is therefore of paramount importance. Another pressing issue is to clarify whether target mass corrections do not
invalidate QCD factorization for coherent DVCS on nuclei \cite{CLAS:2017udk,CLAS:2021ovm}.

An intuitive way to understand the meaning and importance of kinematic power corrections is the following~\cite{Braun:2014paa}. The
leading-twist approximation in DVCS is intrinsically ambiguous since the four-momenta of the initial and final photons and protons do not
lie in one plane. Hence the distinction of longitudinal and transverse directions is convention-dependent. In the Bjorken high-energy limit
this is the  $1/Q$ effect. The freedom to redefine large ``plus'' parton momenta by adding smaller transverse components has two
consequences. First, the relation  of the skewness parameter $\xi$ with  the Bjorken variable $x_B$ may involve power suppressed
contributions. Second, such a redefinition generally leads to excitation of the subleading photon helicity-flip
amplitudes~\cite{Braun:2014sta,Braun:2014paa}. This convention-dependence should be viewed as a theoretical uncertainty and is numerically
rather large, see~\cite{Guo:2021gru} for a detailed study.

At the present time, the kinematic power corrections to DVCS are known to the twist-four accuracy, i.e. up to terms $\sim t/Q^2$ and $\sim
m^2/Q^2$~\cite{Braun:2014sta}.
A typical size of these corrections is of order 10\% for asymmetries, but they can be as large as 100\% for the cross section in certain
kinematics. These corrections can significantly impact the extraction of GPDs from  data and have to be taken into
account~\cite{Defurne:2015kxq,Defurne:2017paw}. The formalism of Refs.~\cite{Braun:2012bg,Braun:2014sta} was used in the most recent study
by the JLAB Hall A collaboration \cite{JeffersonLabHallA:2022pnx}. This publication presents the first experimental extraction of all four
helicity-conserving Compton Form Factors (CFFs) of the nucleon as a function of $x_B$, while systematically including higher-twist helicity
flip amplitudes in the kinematic approximation. It is argued that helicity-flip amplitudes contribute to producing a good fit of the cross
section and most importantly to providing realistic uncertainties on the helicity-conserving CFFs. The helicity-conserving contribution
alone overshoots the data at 180 degrees scattering angle, which is then compensated by helicity-flip contributions~\footnote{C. Munoz
Camacho, private communication}.

Our aim is to develop an approach that would allow one to calculate and possibly resum the corrections $\sim (\sqrt{-t}/Q)^k $ and $\sim
(m/Q)^k$ to all powers. On a more formal level, the task can be formulated as follows. Let $\mathcal O^{\mu_1\ldots \mu_N}$ be local
twist-two operators. The matrix elements of these operators define the GPD moments. The  kinematic contributions we are considering here
receive
 contributions of {\it higher-twist} descendants of the twist-two operators, of the type
\begin{align}
   \partial_{\mu_1}\mathcal O^{\mu_1\mu_2\ldots \mu_N}, \qquad
    \partial_{\mu_1}\partial_{\mu_2}\mathcal O^{\mu_1\mu_2\mu_3\ldots \mu_N}, \qquad
\partial^2 \mathcal O^{\mu_1 \ldots \mu_N}, \quad \text{etc.},
\label{intro:1}
\end{align}
where $\partial_\mu$ is a total derivative. The problem is that matrix elements of the first two operators in \eqref{intro:1} (and similar
ones with more derivatives) vanish for on-mass-shell partons. Hence the usual method to calculate the OPE coefficients functions for these
operators --- evaluate  both sides of the OPE on free quarks --- is not applicable. The technique developed
in~\cite{Braun:2011zr,Braun:2011dg,Braun:2012bg,Braun:2012hq} is based on considering instead quark-antiquark-gluon matrix elements and
using symmetry properties of the corresponding renormalization group equations. Unfortunately this approach becomes too unwieldy beyond
twist four.

In Ref.~\cite{Braun:2020zjm} we suggested a different technique based on the conformal field theory (CFT) methods. In a conformal theory,
the coefficients with which the descendant operators enter the OPE are 
completely determined  by the leading-twist contributions that can be obtained by considering forward matrix
elements~\cite{Ferrara:1971vh,Ferrara:1971zy,Ferrara:1973yt}. For QCD, this  means that kinematic corrections to DVCS amplitudes are
unambiguously determined by  DIS coefficient functions. Of course, QCD is not a conformal theory. However, one can consider a modified
theory, QCD in non-integer $d=4-2\epsilon$ space-time dimensions and fine-tune 
the strong coupling $\alpha_s$ to nullify the $\beta$-function (Wilson-Fisher fixed point~\cite{Wilson:1973jj}). This restores the scale
and conformal invariance  of the correlation functions of gauge-invariant operators~\cite{Braun:2018mxm}. Observables calculated in the
four-dimensional and critical QCD differ beyond leading order by terms proportional to the QCD $\beta$ function. Such terms can be
calculated and added, at least in principle~\cite{Braun:2020yib}, while there are no corrections at the tree level.

 The OPE for the product of two conserved vector
currents in a generic CFT was constructed in Ref.~\cite{Braun:2020zjm}.
The  expansion for the product of two scalar currents was originally obtained  in Ref.~\cite{Ferrara:1971vh} in a different form. A simple
representation for the coefficient functions obtained in~\cite{Braun:2020zjm} is well-suited for studies of high-energy scattering in QCD 
(possible applications beyond DVCS include the studies of $t$-channel processes like $\gamma^\ast\gamma \to \pi\pi$, 
see~\cite{Lorce:2022tiq}).

In this work we use this result to calculate the finite-$t$ and target mass corrections to the helicity amplitudes in DVCS on a scalar
target to the next-to-leading power accuracy and the leading order in the strong coupling. Schematically,
\begin{align}
       \mathcal{A}^{++} &\sim 1 + \frac{1}{Q^2} + \frac{1}{Q^4}\,,
\notag\\
       \mathcal{A}^{0+} &\sim \frac{1}{Q} + \frac{1}{Q^3}\,,
\notag\\
       \mathcal{A}^{-+} &\sim \frac{1}{Q^2} + \frac{1}{Q^4} \,,
\label{intro:2}
\end{align}
where $\mathcal{A}^{++}$, $\mathcal{A}^{0+}$ and $\mathcal{A}^{-+}$ are the helicity-conserving, helicity-flip and double-helicity-flip
amplitudes, respectively, in a particular reference frame~\cite{Braun:2012bg}. Precise definitions are given in the text. An extension to
higher powers is straightforward but unlikely to be relevant for phenomenology, so that we do not work out explicit expressions.

In Section 2 we carry out the first part of this program. Namely, we   rewrite  the OPE obtained in Ref.~\cite{Braun:2020zjm} in terms of
the nonlocal light-ray operators. To this end we develop a certain technique which  relies heavily on   the representation theory of
$\mathrm{SU}(1,1)$ group.
The light-ray OPE in Eq.~\eqref{LROPE} presents the final result for this part.

Matrix elements of light-ray operators are defined in terms of the GPDs.  Thus the Fourier transformation of the expression obtained in
Section 2 yields helicity amplitudes for the DVCS on a chosen target.  
This calculation is
described in Section 3. It is straightforward but proves to be very cumbersome. We find that individual contributions contain infrared (IR)
singularities that cancel in the sum to all orders in the power expansion. We also find that the singularities of the coefficient functions
at the kinematic point $x=\xi$ do not become stronger to all powers, so that the collinear factorization is not endangered. We work our
explicit expressions for the kinematic power corrections to the accuracy indicated in Eq.~\eqref{intro:2} and show that target mass
corrections are not enhanced for nuclear targets. Taking into account these corrections removes the frame dependence of the leading-twist
approximation and restores the electromagnetic gauge invariance of the Compton amplitude up to $1/Q^5$ effects. The final Section 4
contains a short numerical study, our conclusions and outlook.  Some more technical details are given in the Appendices.

%%%%%%%%%%%%%%%%%%%%%%%%%%%%%%%%%%%%%%%%%%%%%%%%%%%%%%%%%%%%%%%%%%%%%%%%%%%%%%%%%%%%%%%%%%%%%%%%%%%%%%%%%%%%%%%%%%%%%%%%%%%%%%%%%%%%%%%%%%%%%
\section{Light-ray operator product expansion} \label{sect:LROPE}
%%%%%%%%%%%%%%%%%%%%%%%%%%%%%%%%%%%%%%%%%%%%%%%%%%%%%%%%%%%%%%%%%%%%%%%%%%%%%%%%%%%%%%%%%%%%%%%%%%%%%%%%%%%%%%%%%%%%%%%%%%%%%%%%%%%%%%%%%%%%%
Our starting expression in this paper is the result of Ref.~\cite{Braun:2020zjm} for the
OPE of two electromagnetic currents taking into account contributions of leading-twist
operators and their higher-twist descendants,   
 cf. Eq.~\eqref{intro:1}%
\footnote{We omit axial-vector contributions as they do not contribute to DVCS on scalar targets.}
 \allowdisplaybreaks{
\begin{align}\label{OPE}
\mathrm{T}\{j^{\mu}(x_1)j^\nu(x_2)\}&=
\frac{1}{i\pi^2}\!\!\sum_{N>0,\text{even}} \! \frac{\rho_N}{N\!+\!1}\int_0^1 \!du\, (u\bar u)^{N}\Biggl\{ \frac{1}{(-x_{12}^2+i0)^2}
\Biggl[
(N+1) g^{\mu\nu}\biggl(1-\frac14 \frac{u\bar u}{N+1} {x_{12}^2}\partial^2\biggr)
\notag\\
&\quad
+\frac1{2N} x_{12}^2\big(\partial_1^\mu\partial_2^\nu-\partial_1^\nu\partial_2^\mu\big)
+\left(1- \frac{u\bar u}{N} \frac{ x_{12}^2\partial^2}{4}\right)\left(
\frac{\bar u}u x_{21}^\mu\partial_1^\nu + \frac{u}{\bar u} x_{12}^\nu\partial_2^\mu\right)
\notag\\
&\quad
- \frac{u\bar u}{N(N\!+\!1)} \frac{x_{12}^2\partial^2}{4} \Big(
 x_{21}^\nu\partial_1^\mu\!+\!  x_{12}^\mu\partial_2^\nu\Big)
-\frac{ x_{12}^\mu x_{12}^\nu}{N\!+\!1}  u\bar u\partial^2\biggl(\!1 - \frac{u\bar u}{N\!+\!2} \frac{x_{12}^2 \partial^2}{4}\biggr)
\Biggr] \mathcal{O}^{(0)}_N\!(x_{21}^u)
\notag\\
&\quad
-\frac1{(-x_{12}^2+i0)}\Biggl[-\frac14 N (\bar u -u)\,g^{\mu\nu}
- \frac{\bar u-u}{4(N+1)} \big( x_{21}^\nu\partial_1^\mu + x_{12}^\mu\partial_2^\nu \big)
\notag\\
&\quad
+\frac12\Big(\bar u\,x_{21}^\mu\partial_1^\nu-u\,x_{12}^\nu\partial_2^\mu  \Big)
+\frac{N}{2(N+2)(N-1)}\Big(x_{21}^\nu\partial_1^\mu - x_{12}^\mu\partial_2^\nu \Big)
\notag\\
&\quad
+\frac14\frac{N(N^2+N+2)}{(N+1)(N+2)(N-1)} \left(\frac{u}{\bar u} x_{12}^\nu\partial_2^\mu
- \frac{\bar u}u x_{21}^\mu\partial_1^\nu \right)
\notag\\
&\quad
+\frac{x_{12}^\mu x_{12}^\nu}{(-x_{12}^2+i0)} (\bar u-u) \frac{N}{N+1}
\left(1 - \frac12 \frac{u\bar u}{N+2} x_{12}^2 \partial^2 \right)
 \Biggr]\,\mathcal{O}^{(1)}_N(x_{21}^u)
 \\
&\quad
- \frac{x_{12}^\mu x_{12}^\nu}{(-x_{12}^2+i0)}\!
 \left[\frac{N^2+N+2}{4(N+1)(N+2)}\! -  \!\frac{u\bar u N(N-1)}{(N+1)(N+2)}
\right] \mathcal{O}_N^{(2)}(x_{21}^u)
\Biggl\}\,+\ldots,\notag
\end{align}
}
where
\begin{align}
\rho_N & =  i^{N-1} \frac{(2N+1)!}{(N-1)!N!\,N!}\,,
\label{rhoN}
\end{align}
\begin{align}
 & \bar u = 1-u\,,\qquad x_{12} = x_1 - x_2\,, \qquad x_{21}^u = \bar u x_2 + u x_1\,,
\qquad \partial_1^\mu = \frac{\partial}{\partial x_1^\mu}\,,\qquad \partial_2^\mu = \frac{\partial}{\partial x_2^\mu}
\end{align}
and a derivative without a subscript $1,2$ stands for 
\begin{align}
\partial^\mu  \mathcal{O}_N^{(k)}(y) = \frac{\partial}{\partial y_\mu} \mathcal{O}_N^{(k)}(y) \,.
\end{align}
For a generic hadronic matrix element between states with different momenta
\begin{align}
 \langle p'|\partial^\mu \mathcal{O}_N^{(k)}(y)|p\rangle  = i \Delta^\mu  \langle p'| \mathcal{O}_N^{(k)}(y)|p\rangle\,,
 \qquad \Delta^\mu = (p'-p)^\mu\,,
\end{align}
so that in what follows 
we will often replace $\partial^\mu \mapsto i \Delta^\mu$,  $\partial^2 \mapsto - \Delta^2$ already on the operator level.

The operators $\mathcal{O}^{(k)}_N$ are defined as
\begin{align}\label{O012}
\mathcal{O}^{(0)}_N(y) &=x_{12,\mu_1}\cdots x_{12,\mu_N}\mathcal O_N^{\mu_1\ldots\mu_N}(y)\,,
\notag\\
\mathcal{O}^{(1)}_N(y) &=x_{12,\mu_2}\cdots x_{12,\mu_N}\frac{\partial}{\partial y^{\mu_1}}\mathcal O_N^{\mu_1\ldots\mu_N}(y)\,,
\notag\\
\mathcal{O}^{(2)}_N(y) &=x_{12,\mu_3}\cdots x_{12,\mu_N}\frac{\partial}{\partial y^{\mu_1}}
\frac{\partial}{\partial y^{\mu_2}}\mathcal O_N^{\mu_1\ldots\mu_N}(y)\,,
\end{align}
where $\mathcal O_N^{\mu_1\ldots\mu_N}$ are multiplicatively renormalizable 
leading-twist operators with spin $N$ normalized
as
\begin{align}
   \mathcal O_N^{\mu_1\ldots\mu_N}(0) = i^{N-1}\bar q(0) \gamma^{\{\mu_1} D^{\mu_2}\ldots D^{\mu_N\}} q(0)
   +\,\text{total~derivatives}\,.
\label{eq:normalization}
\end{align}
Here $\{\ldots\}$ denotes symmetrization and trace subtraction for all enclosed Lorentz indices. In what follows we will use the notation
$[\ldots]_{lt}$ for the leading-twist part of an operator, e.g.,
\begin{align}
  \big[\bar q(0) \gamma^{\mu_1} D^{\mu_2}\ldots D^{\mu_N} q(0)\big]_{lt} = \bar q(0) \gamma^{\{\mu_1} D^{\mu_2}\ldots D^{\mu_N\}} q(0)\,.
\end{align}

In the accepted normalization
\begin{align}
n_{\mu_1}\ldots n_{\mu_N} \mathcal{O}_N^{\mu_1\ldots\mu_N}(y) =
 \frac{\Gamma(3/2) \Gamma(N)}{\Gamma(N+1/2)} \left(\frac{i\partial_+}{{4}}\right)^{N-1}\bar q(y) \gamma_+
C_{N-1}^{3/2}\left(\tfrac{\stackrel{\rightarrow}{D}_+ - \stackrel{\leftarrow}{D}_+}
                      {\stackrel{\rightarrow}{D}_+ + \stackrel{\leftarrow}{D}_+}\right) q(y)\,,
\label{Gegenbauer}
\end{align}
where $n^\mu$ is an arbitrary light-like vector, $n^2=0$,  $D_+ = D^\mu n_\mu$, etc. The expression in Eq.~\eqref{OPE} satisfies exact
electromagnetic Ward identities
\begin{align}
 \partial_1^\mu \mathrm{T}\{j^{\mu}(x_1)j^\nu(x_2)\} =  \partial_2^\nu \mathrm{T}\{j^{\mu}(x_1)j^\nu(x_2)\} =0
\label{Ward}
\end{align}
up to, possibly,  polynomials in $x_{12}^2$ which give rise to delta-function terms  after Fourier transform to the momentum space. The OPE
in this form is term-by term  translation invariant (cf. a discussion in ~\cite{Braun:2011zr,Braun:2011dg})
\begin{align}
  \langle p'| \mathrm{T}\{j^{\mu}(x_1+y)j^\nu(x_2+y)\} |p\rangle &= e^{i(\Delta\cdot y)}
    \langle p'| \mathrm{T}\{j^{\mu}(x_1)j^\nu(x_2)\} |p\rangle\,,
\label{translation}
\end{align}
so that without loss of generality one can make a specific choice, e.g., consider  $\mathrm{T}\{j^{\mu}(x)j^\nu(0)\}$ or
$\mathrm{T}\{j^{\mu}(0)j^\nu(-x)\}$ to simplify the algebra.

%%%%%%%%%%%%%%%%%%%%%%%%%%%%%%%%%%%%%%%%%%%%%%%%%%%%%%%%%%%%%%%%%%%%%%%%%%%%%%%%%%%%%%%%%%%%%%%%%%%%%%%%%%%%%%%%%%%%%%%%%%%%%%%%%%%%%%%%%%%%%
\subsection{Twist expansion} \label{sect:twist}
%%%%%%%%%%%%%%%%%%%%%%%%%%%%%%%%%%%%%%%%%%%%%%%%%%%%%%%%%%%%%%%%%%%%%%%%%%%%%%%%%%%%%%%%%%%%%%%%%%%%%%%%%%%%%%%%%%%%%%%%%%%%%%%%%%%%%%%%%%%%%

The conformal OPE in \eqref{OPE} involves leading-twist operators integrated with a certain weight function over their position on the
straight line connecting the electromagnetic currents. Since the separation $x_{12}$ is not light-like, $x_{12}^2 \slashed{=} 0$, this
integration upsets the twist expansion. Indeed, expanding $\mathcal{O}_N^{(0)}(x_{21}^u)$, e.g., around the middle point
$x_{21}^{\scriptscriptstyle u =1/2} = x^+=\frac12(x_1+x_2)$ one obtains local operators of the form
\begin{align}
 x_{12}^{\nu_1}\ldots  x_{12}^{\nu_k}\,  x_{12}^{\mu_1}\ldots x_{12}^{\mu_N} \partial_{\nu_1}\ldots\partial_{\nu_k}
\bar q(x^+) \gamma_{\{\mu_1} D_{\mu_2}\ldots D_{\mu_N\}} q(x^+)\,,
\end{align}
where not all traces are subtracted. As the first step, we need to rewrite \eqref{OPE} in terms of the leading twist operators
\begin{align}
  \big[\partial_{\nu_1}\ldots\partial_{\nu_k} \bar q  \gamma_{\mu_1} D_{\mu_2}\ldots D_{\mu_N} q\big]_{lt}
 = \partial_{\{\nu_1}\ldots\partial_{\nu_k} \bar q  \gamma_{\mu_1} D_{\mu_2}\ldots D_{\mu_N\}} q\,.
\end{align}
This can be done retaining the structure of the conformal OPE using the technique of Refs.~\cite{Balitsky:1987bk,Balitsky:1990ck}.

For simplicity, take $x_1=x$, $x_2=0$ so that $x_{21}^u = ux$. The leading twist projection of a function $f(x)$ 
satisfies the Laplace equation, $\partial_x^2 [f(x)]_{lt}=0 $, with the boundary condition $[f(x)]_{lt} = f(x)$ at $x^2=0$. The solution
can be written as an expansion in powers of the deviation from the light cone~\cite{Balitsky:1987bk}
\begin{align}
 [f(x)]_{lt} &= f(x) - \frac14 x^2 \int_0^1\frac{dt}{t} \partial_x^2 f(tx) +
 \frac{1}{32} x^4 \int_0^1\frac{dt}{t} \frac{\bar t}{t} \partial_x^4 f(tx) +
\mathcal{O}(x^6)\,.
\label{LTfunction1}
\end{align}
The inverse relation reads
\begin{align}
 f(x) &=  [f(x)]_{lt} + \frac14 x^2 \int_0^1\frac{dt}{t} [\partial_x^2 f(tx)]_{lt} +
 \frac1{32} x^4  \int_0^1 dt\,\frac{\bar t}{t^3} [\partial_x^4\, f(tx)]_{lt}
+ \mathcal{O}(x^6)\,.
\end{align}
Replacing $f(x)$ by $\mathcal{O}_N^{(0)}(ux)$ one obtains
\begin{align}
\mathcal{O}^{(0)}_{N}(u x)  &= [\mathcal{O}^{(0)}_{N}(u x)]_{lt}
 + \frac{x^2}{4}\! \int_0^1\!\frac{dt}{t} \Big[\partial_x^2 e^{i u t \Delta x } t^N \mathcal{O}^{(0)}_{N}( 0) \Big]_{lt}
 +   \frac{x^4}{32}\! \int_0^1\!\frac{\bar t\,dt }{t^3} \partial_x^4 \Big[ e^{i u t \Delta x } t^N \mathcal{O}^{(0)}_{N}( 0) \Big]
\!+ \mathcal{O}(x^6)
\notag\\&=
[\mathcal{O}^{(0)}_{N}(u x)]_{lt}
- \frac{x^2}{4}  \int_0^1\frac{dt}{t} t^N \Big[ u^2 t^2 \Delta^2 [\mathcal{O}^{(0)}_{N}]_{lt}(u t x)
- 2  u t N [\mathcal{O}^{(1)}_{N}]_{lt}(utx) \Big]
\notag\\&\quad
+   \frac{x^4}{32} \int_0^1\!dt\, \frac{\bar t}{t^3} t^N
\biggl\{ u^4 t^4 \Delta^4 [\mathcal{O}^{(0)}_{N}(u t x)]_{lt}
- 4 N u^3 t^3 \Delta^2 [\mathcal{O}^{(1)}_{N}(u t x)]_{lt}
\notag\\&\quad
+ 4 N(N-1) u^2 t^2  [\mathcal{O}^{(2)}_{N}(u t x)]_{lt}
\biggr\}
+ \mathcal{O}(x^6)\,,
\label{O0-lt}
\end{align}
where  taking the matrix element $\langle p'|\ldots |p\rangle$ is tacitly assumed hence {$(p'-p)^\mu = \Delta^\mu \Leftrightarrow -i\partial^\mu$}, and we used that
{$\partial_x^2\mathcal{O}^{(0)}_{N}(0) =0$ because ${\cal O}_N^{\mu_1\cdots\mu_N}(y)$ is a traceless operator, see~\eqref{O012}}. Substituting this expansion in \eqref{OPE} one finds that the $t$-integration can in  most cases
be taken easily so that, e.g.,
\begin{align}\label{OOO}
\int_0^1 du\, (u\bar u)^N
\mathcal {O}^{(0)}_N(ux) &=
\int_0^1 du\, (u\bar u)^N [\mathcal {O}^{(0)}_N(ux)]_{lt} -
\frac{x^2 \Delta^2}{4}\frac{1}{N+1}  \int_0^1 du\,(u\bar u)^{N+1}[\mathcal O_N^{(0)}(ux)]_{lt}
\notag\\ &\quad
+\frac{x^2}{2}\frac N{(N+1)}  \int_0^1 du\, u^N \bar u^{N+1} [\mathcal O_N^{(1)}(ux)]_{lt}
+\ldots\, .
\end{align}
In the similar manner one obtains
\begin{align}
\mathcal{O}^{(1)}_{N}(u x)  &=
[\mathcal{O}^{(1)}_{N}(u x)]_{lt}
- \frac{x^2}{4} \! \int_0^1\frac{dt}{t} t^{N-1} \Big[u^2 t^2 \Delta^2 [\mathcal{O}^{(1)}_{N}]_{lt}(u t x)
- 2 u t (N\!-\!1)[\mathcal{O}^{(2)}_{N}]_{lt}(utx) \Big]+ \mathcal{O}(x^4),
\notag\\
\mathcal{O}^{(2)}_{N}(u x) & =  [\mathcal{O}^{(2)}_{N}(u x)]_{lt} + \mathcal{O}(x^2).
\end{align}
This accuracy is sufficient since the omitted terms only give rise to polynomials in $x^2$ in the OPE and can all be neglected.

%%%%%%%%%%%%%%%%%%%%%%%%%%%%%%%%%%%%%%%%%%%%%%%%%%%%%%%%%%%%%%%%%%%%%%%%%%%%%%%%%%%%%%%%%%%%%%%%%%%%%%%%%%%%%%%%%%%%%%%%%%%%%%%%%%%%%%%%%%%%%
\subsection{Light-ray operator representation} \label{sect:lt}
%%%%%%%%%%%%%%%%%%%%%%%%%%%%%%%%%%%%%%%%%%%%%%%%%%%%%%%%%%%%%%%%%%%%%%%%%%%%%%%%%%%%%%%%%%%%%%%%%%%%%%%%%%%%%%%%%%%%%%%%%%%%%%%%%%%%%%%%%%%%%

%%%%%%%%%%%%%%%%%%%%%%%%%%%%%%%%%%%%%%%%%%%%%%%%%%%%%%%%%%%%%%%%%%%%%%%%%%%%%%%%%%%%%%%%%%%%%%%%%%%%%%%%%%%%%%%%%%%%%%%%%%%%%%%%%%%%%%%%%%%%%
\subsubsection{Methods}
%%%%%%%%%%%%%%%%%%%%%%%%%%%%%%%%%%%%%%%%%%%%%%%%%%%%%%%%%%%%%%%%%%%%%%%%%%%%%%%%%%%%%%%%%%%%%%%%%%%%%%%%%%%%%%%%%%%%%%%%%%%%%%%%%%%%%%%%%%%%%

The next step is to rewrite the answer in terms of nonlocal light-ray operators
\begin{align}
  \mathscr{O}(z_1,z_2) = \frac12 \Big[\bar q (z_1 x)\slashed{x} [z_1 x, z_2 x] q(z_2 x)
    - \bar q (z_2 x)\slashed{x}[z_2 x, z_1 x] q(z_1 x)\Big]_{lt},
\label{LRoperator}
\end{align}
where $z_1, z_2$ are real numbers, $[z_1 n, z_2 n]$ is the Wilson line, and the nonlocal quark-antiquark operators on the r.h.s. are
understood as generating functions for renormalized leading--twist local operators. This representation is advantageous since the matrix
elements of light-ray operators are expressed directly in terms of GPDs. It will allow us to calculate power corrections to the
DVCS helicity amplitudes (Compton form factors) directly, bypassing the nontrivial problem of analytic continuation from the set of moments (matrix
elements of local operators). In this section we derive the light-ray operator representation for  $\mathrm{T}\{j^{\mu}(x)j^\nu(0)\}$, i.e.
we set $x_1=x$, $x_2=0$ that results in some simplifications.

The expansion of the light-ray operator \eqref{LRoperator} over the  local operators \eqref{O012}, \eqref{Gegenbauer}  reads \cite{Braun:2011zr}
\begin{align}\label{relation-zero}
\mathscr{O}(z_1,z_2) &=  \sum_{\substack{N>0,\\\text{even}}}
\rho_N z_{12}^{N-1}\,\int_0^1 du\, (u\bar u)^{N}\big[\mathcal {O}^{(0)}_N( z_{21}^ux )\big]_{lt},
\end{align}
where the coefficients $\rho_N$ are defined in \eqref{rhoN}. The leading contribution $\sim g_{\mu\nu}/x_{12}^4$ in the first line in
Eq.~\eqref{OPE} has exactly this form, so that it can be readily  written in terms of $\mathscr{O}(1,0)$ (for $x_1=x$, $x_2=0$). A generic
contribution to the OPE has the form
\begin{align}
 \sum_{\substack{N>0,\\\text{even}}}  \rho_N  f(N) \,\int_0^1 du\, (u\bar u)^{N} g(u) \big[\mathcal {O}^{(k)}_N(u x )\big]_{lt},
\label{generic}
\end{align}
and the task is to rewrite such expressions as certain integrals of light-lay operators. For example,
\begin{align}
  \sum_{\substack{N>0,\\\text{even}}}  \rho_N \frac{1}{N+1} \,\int_0^1 du\, (u\bar u)^{N} \frac{u}{\bar u} \big[\mathcal {O}^{(0)}_N(ux)\big]_{lt}
= \int_0^1\!dv\, \mathscr{O}(1,v)\,.
\end{align}
This relation can be easily verified  using \eqref{relation-zero}  and performing one integration.

For a certain class of functions, the necessary expressions can be worked out using conformal symmetry.
The expression in Eq.~\eqref{relation-zero} can equivalently be rewritten as \cite{Braun:2011dg}{\footnote{Notice the difference in the definition of $N$.}}
\begin{align}
 \mathscr{O}(z_1,z_2) &= \sum_{\substack{N>0,\\\text{even}}}^\infty \sum_{k=0}^\infty \omega_{Nk} (S_+^{(1,1)})^k z_{12}^{N-1}
{\big[(x\partial)^k \mathcal{O}_N^{(0)}(0)\big]_{lt}},
\label{relation-0}
\end{align}
where
\begin{align}
 \omega_{Nk} &= \frac{\rho_N}{k!} \frac{\Gamma(N+1)\Gamma(N+1)}{\Gamma(2N+2+k)}
\label{omegaNk}
\end{align}
and $S_{+}^{(j_1,j_2)}$ is one of the generators of the $\mathrm{SL}(2,\mathrm R)$ group (a collinear subgroup of conformal transformations
\cite{Braun:2003rp})
\begin{align}
 S_-^{(j_1,j_2)} &= - \partial_{z_1} - \partial_{z_2}\,,
\notag\\
 S_0^{(j_1,j_2)} &=z_1\partial_{z_1} + z_2\partial_{z_2} +j_1+j_2,
\notag\\
 S_+^{(j_1,j_2)}&=z^2_1\partial_{z_1} + z^2_2\partial_{z_2} +2j_1 z_1+2j_2 z_2.
\label{SL2generators}
\end{align}
Here $j_k$ (conformal spins) specify the irreducible representation of the $\mathrm{SL}(2,\mathrm R)$ group $T^{(j_k)}$ \cite{MR3469700}.
The operators in \eqref{SL2generators} act on the tensor product $T^{(j_1)}\otimes T^{(j_2)}$.

Let $\mathrm{H}$ be an $\mathrm{SL}(2,\mathrm R)$-invariant operator acting on field coordinates (i.e., it commutes with the symmetry
generators). 
It can be written in the form%
\footnote{see appendix B in \cite{Braun:2009vc} and references therein}
\begin{align}
  \mathrm{H}\phi(z_1,z_2) &= \int_0^1 \!d\alpha\! \int_0^{\bar\alpha}\!\! d\beta\, h(\tau) \phi(z_{12}^\alpha,z_{21}^\beta)\,,
\qquad \tau = \frac{\alpha\beta}{\bar\alpha\bar\beta}\,.
\label{11->11}
\end{align}
Translation-invariant polynomials $z_{12}^k$ are eigenfunctions of any invariant operator, and the weight function (kernel) $h(\tau)$ is
uniquely determined by its spectrum
\begin{align}
  \mathrm{H} z_{12}^{N-1} &= h_N  z_{12}^{N-1} = z_{12}^{N-1} \int_0^1 \!d\alpha\! \int_0^{\bar\alpha}\!\! d\beta\, h(\tau) (1-\alpha-\beta)^{N-1}\,.
\label{SL2-1}
\end{align}
If $h_N$ satisfies the so-called reciprocity relation~\cite{Dokshitzer:2005bf,Basso:2006nk,Alday:2015eya,Alday:2015ewa},
$h_N = h_{-N-1}$, finding the corresponding kernel $h(\tau)$ is usually not difficult, e.g.,
\begin{align}
   & h_N = \frac{1}{N(N+1)} && \implies& & h(\tau) =1\,,
\notag\\
   & h_N = \frac{1}{N^2(N+1)^2} &&\implies& & h(\tau) = - \ln\bar\tau\,.
\label{SL2-2}
\end{align}

Applying the invariant operator \eqref{11->11} to Eq.~\eqref{relation-0} one obtains
\begin{align}
 \mathrm{H} \mathscr{O}(z_1,z_2) &= \sum_{\substack{N>0,\\\text{even}}}^\infty \sum_{k=0}^\infty \omega_{Nk} (S_+^{(1,1)})^k h_N z_{12}^{N-1}
\big[(x\partial)^k \mathcal{O}_N^{(0)}(0)\big]_{lt},
\label{relation-0a}
\end{align}
or, going back to the representation in \eqref{relation-zero}
\begin{align}\label{relation-HH}
\sum_{\substack{N>0,\\\text{even}}}\rho_N h_N  z_{12}^{N-1}\,\int_0^1 du\, (u\bar u)^{N}\big[\mathcal {O}^{(0)}_N( z_{21}^ux )\big]_{lt}
& =   \int_0^1 \!d\alpha\! \int_0^{\bar\alpha}\!\! d\beta\, h(\tau)\, \mathscr{O}(z^\alpha_{12},z^\beta_{21})\,.
\end{align}
This relation allows one to derive a light-ray operator representation for the sum in Eq.~\eqref{generic} if $g(u)=1$ in~\eqref{generic} and
$f(N)$ satisfies the reciprocity relation $f(N)= f(-N-1)$.

Other cases can be treated similarly, but the derivation becomes more involved.
One new element is that instead of invariant operators  $\mathrm{H}:\,T^{(1)}\otimes T^{(1)}\mapsto T^{(1)}\otimes T^{(1)}$
which commute with the $S_k^{(1,1)}$ generators, $\mathrm{H}\,S_k^{(1,1)} = S_k^{(1,1)}\mathrm{H} $,
one needs to consider intertwining operators between different
representations,  e.g. $\widetilde{\mathrm{H}}:\,T^{(1)}\otimes T^{(1)}\mapsto T^{(\frac32)}\otimes T^{(\frac12)}$,
such that $\widetilde{\mathrm{H}}\,S_k^{(1,1)} = S_k^{(\frac32,\frac12)}\widetilde{\mathrm{H}}$.
Another issue is that known light-ray operator representations involving $\mathcal {O}^{(1)}_N$ and $\mathcal {O}^{(2)}_N$
are more complicated as compared to \eqref{relation-zero}:
\begin{align}\label{O1genfunction}
\sum_{\substack{N>0\\ \text{even}}} \rho_N N^2 z_{12}^{N-1}\int_0^1\! du\, (u\bar u)^N \Big[\mathcal O_N^{(1)}(z_{21}^u x)\Big]_{lt}
&=
 \left(S_0^{(1,1)}-1\right) (i\Delta\partial_x)\mathscr{O}(z_1,z_2) + \frac12 S_{+}^{(1,1)} \Delta^2 \mathscr{O}(z_1,z_2)\, ,
\end{align}
and
\begin{multline}\label{O2genfunction}
\sum_{\substack{N>0\\ \text{even}}}
\rho_N N^2 \int_0^1\! du\, (u\bar u)^N \left\{
(N-1)^2 \big[\mathcal O_N^{(2)}(z_{21}^u x)\big]_{lt}
+
\Delta^2S_+^{(1,1)} \int_0^1 dt t^{2N+1} \big[\mathcal O_N^{(1)}(t z_{21}^u x)\big]_{lt}\right\} z_{12}^{N-1}
\\
=
\left\{ \big(S_0^{(1,1)}-2\big) (i\Delta\partial_x) + \frac12 \Delta^2 S_+^{(1,1)}\right\}
\left\{\big(S_0^{(1,1)}-1\big) (i\Delta\partial_x) + \frac12 \Delta^2S_+^{(1,1)}\right\}
\mathscr{O}(z_1,z_2).
\end{multline}
These relations can be obtained following the technique of Ref.~\cite{Braun:2011dg}, see appendix~\ref{app:O2}.

%%%%%%%%%%%%%%%%%%%%%%%%%%%%%%%%%%%%%%%%%%%%%%%%%%%%%%%%%%%%%%%%%%%%%%%%%%%%%%%%%%%%%%%%%%%%%%%%%%%%%%%%%%%%%%%%%%%%%%%%%%%%%%%%%%%%%%%%%%%%%
\subsubsection{Example}\label{sect:Example}
%%%%%%%%%%%%%%%%%%%%%%%%%%%%%%%%%%%%%%%%%%%%%%%%%%%%%%%%%%%%%%%%%%%%%%%%%%%%%%%%%%%%%%%%%%%%%%%%%%%%%%%%%%%%%%%%%%%%%%%%%%%%%%%%%%%%%%%%%%%%%

To demonstrate how it works, consider terms $\sim g_{\mu\nu}$ in the OPE \eqref{OPE}
\begin{align}\label{OPEgmunu}
\mathrm{T}\{j^{\mu}(x)j^\nu(0)\}&=
\frac{g^{\mu\nu}/i\pi^2}{(-x^2+i0)^2}\!\!\sum_{\substack{N>0\\ \text{even}}} \! \frac{\rho_N}{N\!+\!1}\int_0^1 \!du\, (u\bar u)^{N}
\biggl\{
(N+1) \biggl(1+ \frac14 \frac{u\bar u}{N+1} {x^2}\Delta^2\biggr)\mathcal{O}^{(0)}_N\!(u x )
\notag\\
&\quad
-\frac14 x^2  N (\bar u -u)\,\mathcal{O}^{(1)}_N(u x)
\biggl\}\,+\ldots,
\end{align}
where we replaced $\partial^2\mapsto - \Delta^2$. The ellipses stand for the other existing Lorentz structures.

At the first step we use Eq.~\eqref{OOO} to rewrite the most singular $1/x^4$ contributions in terms of the leading-twist operators,
\begin{align}
\int_0^1 du\, (u\bar u)^N
\mathcal {O}^{(0)}_N(ux) &=
\int_0^1 du\, (u\bar u)^N [\mathcal {O}^{(0)}_N(ux)]_{lt} -
\frac{x^2 \Delta^2}{4}\frac{1}{N+1}  \int_0^1 du\,(u\bar u)^{N+1}[\mathcal O_N^{(0)}(ux)]_{lt}
\notag\\ &\quad
+\frac{x^2}{2}\frac N{(N+1)}  \int_0^1 du\, u^N \bar u^{N+1} [\mathcal O_N^{(1)}(ux)]_{lt}
+\ldots
\label{repeat1}
\end{align}
In all other contributions one can simply replace $\mathcal O_N^{(k)}$ by $[\mathcal O_N^{(k)}]_{lt}$ to the required accuracy. The second
term on the r.h.s. of Eq.~\eqref{repeat1} (the term $\sim \Delta^2$) cancels against the corresponding contribution in \eqref{OPEgmunu}.
Adding together the two terms  $\sim \mathcal O_N^{(1)}$ one gets
\begin{flalign}\label{TjjT1}
\mathrm{T}\{j^{\mu}(x)j^\nu(0)\}&= \frac{ g_{\mu\nu}/(i\pi^2)}{(-x^2+i0)^2}\biggl\{
\mathscr{O}(1,0) + \frac{x^2}4  \sum_{\substack{N>0\\ \text{even}}} \rho_N \frac N{N+1}\int_0^1 \!du\, (u\bar u)^{N} \mathcal O_N^{(1)}(ux)
\biggr\} + \ldots\,,&
\end{flalign}
where we used \eqref{relation-zero} to rewrite the leading contribution in terms of the light-ray operator.

The next step is to make use of the identity \eqref{O1genfunction}. 
The sum in~\eqref{TjjT1} differs from that in~\eqref{O1genfunction} by the factor $1/(N(N+1))$ which can be emulated by the application of
the $\mathrm{SL}(2,\mathrm R)$-invariant operator $\mathcal H_+:\,T^{(1)}\otimes T^{(1)}\mapsto T^{(1)}\otimes T^{(1)}$, cf.
Eqs.~\eqref{11->11}, \eqref{SL2-2}:
\begin{align}
\mathcal H_+ f(z_1,z_2) & = \int_0^1d\alpha\int_0^{\bar\alpha} d\beta \, f(z_{12}^\alpha,z_{21}^\beta)\, .
\end{align}
Thus we get
\begin{multline}\label{O1genfunction-1}
\sum_{\substack{N>0\\ \text{even}}}\rho_N
\frac{N}{N+1}  z_{12}^{N-1}\int du (u\bar u)^N \Big[\mathcal O_N^{(1)}(z_{21}^u x)\Big]_{lt}
=
\\ =
    \left(S_0^{(1,1)}-1\right) \mathcal H_+ (i\Delta\partial_x) \mathscr O_+(z_1,z_2)
    + \frac12 S_{+}^{(1,1)} \mathcal H_+ \Delta^2 \mathscr O_+(z_1,z_2)\,,
\end{multline}
where we used that $[S^{(1,1)}_\alpha,\mathcal H_+]=0$.

We need the r.h.s. of Eq.~\eqref{O1genfunction-1} for $z_1=1$, $z_2=0$. In this case one can replace,
\begin{align}
S_+^{(1,1)}\mapsto \mathcal{S}= z_{12}^{-1} \partial_1 z_{12}^2, && S_0^{(1,1)}\mapsto \mathcal{S}=z_{12}^{-1} \partial_1 z_{12}^2.
\label{1->3/2}
\end{align}
The operator $\mathcal{S}$ is an invariant operator with eigenvalues $N+1$ \footnote{Indeed, $\mathcal{S} z_{12}^{N-1} = (N+1)\,
z_{12}^{N-1}$.} which intertwines the representations of the $\mathrm{SL}(2,\mathrm R)$ group: $\mathcal{S}:T^{(1)}\otimes T^{(1)}\mapsto
T^{(3/2)}\otimes T^{(1/2)}$. Thus the product $\mathcal{S} \mathcal H_+$ is also an invariant operator $T^{(1)}\otimes T^{(1)}\mapsto
T^{(3/2)}\otimes T^{(1/2)}$ with the eigenvalues $(N+1)\times 1/(N(N+1)) = 1/N$ using~\eqref{SL2-1}. Any such operator  can be written in  the form (cf.
\eqref{11->11})
\begin{align}
 \mathcal{S} \mathcal H_+ f(z_1,z_2) &= \int_0^1 \!d\alpha\! \int_0^{\bar\alpha}\!\! d\beta\, \frac{\beta}{\bar\beta}
     w(\tau) \phi(z_{12}^\alpha,z_{21}^\beta)\,,
\qquad \tau = \frac{\alpha\beta}{\bar\alpha\bar\beta}\,,
\end{align}
where the kernel $w(\tau)$ is  uniquely determined by the spectrum. In the case under consideration
\begin{align}
 \int_0^1 \!d\alpha\! \int_0^{\bar\alpha}\!\! d\beta\, \frac{\beta}{\bar\beta}   w(\tau) (1-\alpha-\beta)^{N-1} = \frac{1}{N}
\qquad\Rightarrow\qquad  w(\tau) = \delta(\tau)\,.
\end{align}
Thus
\begin{align}
 \mathcal{S} \mathcal H_+ f(z_1,z_2)= \int_0^1 d\beta \, f(z_1,z_{21}^\beta)\,.
\end{align}
Collecting everything, we obtain the desired representation
\begin{align}
\mathrm{T}\{j^{\mu}(x)j^\nu(0)\}& =
\frac{ g_{\mu\nu}/(i\pi^2)}{(-x^2+i0)^2}\biggl\{ \mathscr{O}(1,0) -
\frac{ x^2}{4}
\int_0^1\!d\alpha\!\int_0^{\bar\alpha} \! d\beta\, (i\Delta\partial_x) \mathscr{O}(\bar\alpha,\beta)
\notag\\
&\quad
+ \frac{ x^2}{4} \Big((i\Delta\partial_x) + \frac12\Delta^2\Big)  \int_0^1 d\beta\, \mathscr{O}(1,\beta)
\biggr\} + \ldots\,.
\end{align}
%

%%%%%%%%%%%%%%%%%%%%%%%%%%%%%%%%%%%%%%%%%%%%%%%%%%%%%%%%%%%%%%%%%%%%%%%%%%%%%%%%%%%%%%%%%%%%%%%%%%%%%%%%%%%%%%%%%%%%%%%%%%%%%%%%%%%%%%%%%%%%%
\subsubsection{Result}
%%%%%%%%%%%%%%%%%%%%%%%%%%%%%%%%%%%%%%%%%%%%%%%%%%%%%%%%%%%%%%%%%%%%%%%%%%%%%%%%%%%%%%%%%%%%%%%%%%%%%%%%%%%%%%%%%%%%%%%%%%%%%%%%%%%%%%%%%%%%%

The other terms in \eqref{OPE} can be treated along the similar lines.
Introducing the notations
\begin{align}
\mathscr{O}_1(z_1,z_2) =
(i\Delta\partial_x) \mathscr{O}(z_1,z_2)\,,
&&
\mathscr{O}_2(z_1,z_2) = \Big((i\Delta\partial_x) + \frac12\Delta^2\Big) \mathscr{O}(z_1,z_2)\,,
\end{align}
we obtain the final result as follows:
\begin{align}
\label{LROPE}
 & %\hspace*{-0.1cm}
\mathrm{T}\{j^{\mu}(x)j^\nu(0)\} =
\notag\\
&=  {\frac{1}{i\pi^2}}
\Biggl\{\frac{1}{x^4} \Biggl[
g^{\mu\nu}\mathscr O(1,0)-x^\mu\partial^\nu\int_0^1\!du\,\mathscr O(\bar u,0)
    -x^\nu(\partial^\mu-i\Delta^\mu)\int_0^1 \!dv\,\mathscr O(1,v)\Biggl]
\notag\\
&\quad
 +\frac{1}{x^2} \left[\frac i 2\big( \Delta^\nu\partial^\mu  -\Delta^\mu \partial^\nu\big)
 \int_0^1\!\!du\!\int_0^{\bar u} \!\!\!dv\, \mathscr O(\bar u,v)
-\frac{\Delta^2}4x^\mu\partial^\nu
\int_0^1\!\!du\,u\! \int_0^{ \bar u}\!\! dv\, \mathscr O(\bar u,v)
\right]
\notag\\
&\quad
+\frac{\Delta^2 }2 \frac{x^\mu x^\nu}{x^4}\int_0^1\!\!du\,\bar u\!\int_0^{\bar u}\!\! dv\, \mathscr O(\bar u,v)
%\notag\\
%&\quad
+\frac1{4x^2}{g^{\mu\nu}}
\left[
-\int_0^1\!\!du\!\int_0^{\bar u} \!\!\!dv\, \mathscr O_1(\bar  u, v)
+\int_0^1\!\! d v \,\mathscr O_2(1, v)
\right]
\notag\\
&\quad -\frac{1}{4x^2}
(x^\nu\partial^\mu + x^\mu\partial^\nu-i x^\mu\Delta^\nu)
\int_0^1\!\!du\!\int_0^{\bar u} \!\!\!dv\,
\left(\ln\bar\tau\,\mathscr O_1(\bar  u, v)
+\frac{ v}{\bar  v}\,\mathscr O_2(\bar u, v)
\right)
\notag\\
%B_4
&\quad -\frac{1}{2x^2}
{(x^\nu\partial^\mu-x^\mu \partial^\nu +ix^\mu\Delta^\nu)} 
\int_0^1\!\!du\!\int_0^{\bar u} \!\!\!dv\,
\frac\tau{\bar\tau}\left(-\mathscr O_1(\bar  u, v)
+\frac{\bar u}{ u}\,\mathscr O_2(\bar u, v)
\right)
\notag\\
%B_5
&\quad- \frac{1}{4x^2}
{x^\nu(\partial^\mu-i\Delta^\mu) }\left[
\int_0^1\!\!du\!\int_0^{\bar u} \!\!\!dv\,
\frac v{\bar v}\left[-2\left(1+\frac{2\tau}{\bar\tau}\right)\,\mathscr O_1(\bar u, v)
+\frac{ v}{\bar v}\mathscr O_2(\bar u, v)
\right]
+\int_0^1\!\! d v \frac{v}{\bar v}\mathscr O_2(1, v)
\right]
\notag\\
&\quad -\frac{1}{2x^2} x^\mu\partial^\nu
\int_0^1\!\!du\!\int_0^{\bar u} \!\!\!dv\,
\biggl[(\ln\bar u +  u)\,\mathscr O_1(\bar  u, v)
+ \bar u \,\mathscr O_2(\bar  u, v) - \frac12 \left(1+\frac{4\tau}{\bar\tau}\right) \mathscr O_2(\bar  u, v)\biggr]
\notag\\
&\quad
-\frac{x^\mu x^\nu}{x^4}
\int_0^1\!\!du\!\int_0^{\bar u} \!\!\!dv\,
\left[
    (\ln\bar\tau+\ln\bar u + u)\,\mathscr O_1(\bar  u, v)
+\left( \frac{ v}{\bar  v} + \bar u \right)\,
\mathscr O_2(\bar u, v)
\right]
\notag\\
&\quad
-
\frac{x^\mu x^\nu}{4x^2}
\Big[(i\Delta \partial) +\frac12\Delta^2\Big]
\int_0^1\!\!du\!\int_0^{\bar u} \!\!\!dv\,
\frac{ v}{\bar v}\left( \frac2{\bar\tau}-1\right)
\mathscr O_1(\bar u, v)
\notag\\
&\quad
+
\frac{x^\mu x^\nu}{2x^2}
\Big[(i\Delta\partial) +\frac14\Delta^2\Big]
\int_0^1\!\!du\!\int_0^{\bar u} \!\!\!dv\,
\left(\ln\bar\tau+\frac{2\tau}{\bar\tau}\right)
\mathscr O_1(\bar u, v)\Biggr\},
\end{align}
where $\partial_\mu = \partial/\partial x^\mu$.

This expression is derived from \eqref{OPE} without any approximations so that it satisfies the
Ward identity \eqref{Ward} and the translation invariance relation \eqref{translation}. The latter becomes hidden, however:
it is only valid in the sum of all terms and not easy to check explicitly. We did not find a simple way to obtain
a light-ray operator representation for the general case $\mathrm{T}\{j^{\mu}(x_1)j^\nu(x_2)\}$.  This restriction, however, poses no issues for 
the application which we pursue next.

%%%%%%%%%%%%%%%%%%%%%%%%%%%%%%%%%%%%%%%%%%%%%%%%%%%%%%%%%%%%%%%%%%%%%%%%%%%%%%%%%%%%%%%%%%%%%%%%%%%%%%%%%%%%%%%%%%%%%%%%%%%%%%%%%%%%%%%%%%%%%
\section{Helicity amplitudes}
%%%%%%%%%%%%%%%%%%%%%%%%%%%%%%%%%%%%%%%%%%%%%%%%%%%%%%%%%%%%%%%%%%%%%%%%%%%%%%%%%%%%%%%%%%%%%%%%%%%%%%%%%%%%%%%%%%%%%%%%%%%%%%%%%%%%%%%%%%%%%
%%%%%%%%%%%%%%%%%%%%%%%%%%%%%%%%%%%%%%%%%%%%%%%%%%%%%%%%%%%%%%%%%%%%%%%%%%%%%%%%%%%%%%%%%%%%%%%%%%%%%%%%%%%%%%%%%%%%%%%%%%%%%%%%%%%%%%%%%%%%%
\subsection{Kinematics and notations}
%%%%%%%%%%%%%%%%%%%%%%%%%%%%%%%%%%%%%%%%%%%%%%%%%%%%%%%%%%%%%%%%%%%%%%%%%%%%%%%%%%%%%%%%%%%%%%%%%%%%%%%%%%%%%%%%%%%%%%%%%%%%%%%%%%%%%%%%%%%%%
%%%%%%%%%%%%%%%%%%%%%%%%%%%%%%%%%%%%%%%%%%%%%%%%%%%%%%%%%%%%%%%%%%%%%%%%%%%%%%%%%%%%%%%%%%%%%%%%%%%%%%%%%%%%%%%%%%%%%%%%%%%%%%%%%%%%%%%%%%%%%
\subsubsection{Helicity decomposition of the Compton tensor}
%%%%%%%%%%%%%%%%%%%%%%%%%%%%%%%%%%%%%%%%%%%%%%%%%%%%%%%%%%%%%%%%%%%%%%%%%%%%%%%%%%%%%%%%%%%%%%%%%%%%%%%%%%%%%%%%%%%%%%%%%%%%%%%%%%%%%%%%%%%%%

The hadronic part of the DVCS amplitude is given by the matrix element of the time-ordered product of two electromagnetic currents
\begin{equation}
  j^{\rm em}_\mu(x) = \bar q(x)\gamma_\mu \mathrm{Q}\, q(x)\,,
\label{jem}
\end{equation}
where $q = \{u,d,\ldots\}$ is the quark field and $\mathrm{Q}$ is the diagonal matrix of quark charges
\begin{align}
\mathrm{Q}=e\,\begin{pmatrix}
e_u& 0 & 0\\
0& e_d & 0\\
\vdots & \vdots & \ddots
\end{pmatrix},
&& e=\sqrt{4\pi\alpha}.
\end{align}
Using translation invariance \eqref{translation} one can write the DVCS amplitude as
\begin{align}\label{amplitude1}
\mathcal{A}_{\mu\nu}=i\int  d^4 x\, e^{-i q x}
\langle p'|T\{j^{\rm em}_\mu(x)j^{\rm em}_\nu(0)\}|p\rangle\,,
\end{align}
or, equivalently,
\begin{align}\label{amplitude2}
\mathcal{A}_{\mu\nu}=i\int  d^4 x\, e^{i q' x}
\langle p'|T\{j^{\rm em}_\nu(x)j^{\rm em}_\mu(0)\}|p\rangle\,,
\end{align}
where $q$ and $q'$ are the ingoing (virtual) and outgoing (real) photon momenta, respectively:
\begin{align}
 q^2 = -Q^2, && q'^2 = 0\,.
\end{align}
The representation in \eqref{amplitude2} proves to be more convenient for our purposes as it leads to much simpler Fourier integrals
in the $q'^2\to 0$ limit.

The DVCS amplitude $\mathcal{A}_{\mu\nu}$ can be written in terms of several scalar functions. We will use the decomposition
suggested in Ref.~\cite{Braun:2012bg}:
\begin{align}\label{covf}
\mathcal{A}^{\mu\nu}=&-g^{\mu\nu}_{\perp} \,\mathcal{A}^{(0)}+\frac{1}{\sqrt{-q^2}}\left(q^\mu-{q'}^\mu \frac{q^2}{(qq')}\right)
P_\perp^\nu
\mathcal{A}^{(1)}
+\frac12\left(P_\perp^\mu P_\perp^\nu - \widetilde P_\perp^\mu \widetilde P_\perp^\nu
\right)
\mathcal{A}^{(2)}+q'_\nu\mathcal{A}_\mu^{(3)}\,,
\end{align}
where
\begin{align}
g_{\mu\nu}^\perp = g_{\mu\nu}-\frac{q_\mu q'_\nu+q'_\mu q_\nu}{(qq')}+{q'_\mu}q'_\nu\frac{q^2}{(qq')^2}\,,
&&
\epsilon_{\mu\nu}^\perp = \frac1{(qq')}\epsilon_{\mu\nu\alpha\beta}{q^\alpha q'^\beta}\, ,
\end{align}
and
\begin{align}
 P^\mu = \frac12(p+p')^\mu\,, && P_\perp^\mu =g_\perp^{\mu\nu} P_\nu\,, && \widetilde P_\perp^\mu =\epsilon_\perp^{\mu\nu} P_\nu\,.
\label{Pperp}
\end{align}
In the frame of reference where the two photon momenta are used to define the longitudinal plane (in four dimensions),
one can define the longitudinal $\varepsilon^0_\mu$ and transverse $\varepsilon^\pm_\mu$ photon polarization vectors
\begin{align}
\varepsilon^0_\mu=-\left(q_\mu-q'_\mu {q^2}/{(q\cdot q')}\right)/{\sqrt{-q^2}}\,,
&&
{\varepsilon^\pm_\mu=(P^\perp_\mu\pm i \widetilde P^\perp_\mu)/ {(\sqrt{2}|P_\perp|)}}\,,
\label{varepsilon^a_mu}
\end{align}
where $|P_\perp|= \sqrt{-P_\perp^2}$, and rewrite~\eqref{covf} as
\begin{align}
\mathcal{A}^{\mu\nu}=\varepsilon_\mu^+\varepsilon_\nu^- \mathcal{A}^{++} + \varepsilon_\mu^-\varepsilon_\nu^+\mathcal{A}^{--}
  + \varepsilon_\mu^0\varepsilon_\nu^- \mathcal{A}^{0+} + \varepsilon_\nu^0\varepsilon_\nu ^+ \mathcal{A}^{0-}
+
\varepsilon_\mu^+\varepsilon_\nu^+ \mathcal{A}^{+-} + \varepsilon_\mu^-\varepsilon_\nu^-\mathcal{A}^{-+},
\end{align}
where
\begin{align}\label{HelicityA}
\mathcal A^{\pm\pm}  = \mathcal A^{(0)}, && \mathcal A^{0\pm}  = - \frac{|P_\perp|}{\sqrt 2} \mathcal A^{(1)}, &&
\mathcal A^{\pm\mp} = \frac{|P_\perp|^2}2 \mathcal A^{(2)}.
\end{align}
One sees that the invariant functions  $\mathcal A^{(0)}$ and $\mathcal A^{(2)}$ have the physical meaning of
helicity-conserving and helicity-flip scattering amplitudes of transversely polarized photons respectively, in this frame.
The amplitude $\mathcal A^{(1)}$ corresponds to the contribution of the longitudinally polarized virtual photon in the initial state. 
The amplitude $\mathcal{A}_\mu^{(3)}$ does not contribute to  physical observables. The invariant functions $\mathcal A^{(k)}$ 
alias the helicity amplitudes $\mathcal A^{\pm\pm}$, $\mathcal A^{0\pm}$, $\mathcal A^{\pm\mp}$ can easily be related to, e.g., the Belitsky-M{\"u}ller-Ji (BMJ) Compton form factors~\cite{Belitsky:2012ch} as described in detail in Ref.~\cite{Braun:2014sta}.

The momentum transfer to the target in this frame 
    is, by construction, purely longitudinal
\begin{align}
  \Delta = p'-p= q-q', && t=\Delta^2, && g_{\mu\nu}^\perp\Delta^\nu =0
\end{align}
and the (space-like) vector $P_\perp^\mu$ has a meaning of the transverse momentum of the target,
which is the same before and after the collision,
\begin{align}
 P_\perp^2 = - |P_\perp^2| = m^2 \frac{t_{\mathrm min} - t}{t_{\mathrm min}}\,.
\end{align}
Here $m$ is the target mass and $t_{\mathrm min} < 0 $ is the smallest kinematically allowed invariant momentum transfer
\begin{align}
 t_{\mathrm min} = - \frac{4 \xi^2 m^2}{1-\xi^2},  && \text{ or } &&  \xi \le \xi_{\text{max}}= \frac1{\sqrt{1- 4m^2/t}},
\label{tmin}
\end{align}
where $\xi$ is the asymmetry (skewness) parameter that we define with respect to the projection on the photon momentum in the final state,
$p_+ = p^\mu q'_\mu$,
\begin{align}
\xi & =\frac{p_+-p'_+}{p_++p'_+} = \frac{x_B(1+t/Q^2)}{2-x_B(1-t/Q^2)}\,, \qquad x_B = \frac{Q^2}{2pq}\,.
\label{xi}
\end{align}
Different helicity amplitudes can be separated using  projection operators:
\begin{align}
   \Pi^{(0)}_{\mu\nu} = P_\mu^\perp P_\nu^\perp +   \widetilde{P}^\perp_\mu  \widetilde{P}^\perp_\nu\,,
&&
   \Pi^{(1)}_{\mu\nu} = q'_\mu P_\nu^\perp \,,
&&
   \Pi^{(2)}_{\mu\nu} = P_\mu^\perp P_\nu^\perp -   \widetilde{P}^\perp_\mu  \widetilde{P}^\perp_\nu\,,
\label{projectors}
\end{align}
so that
\begin{align}
\notag\\
   \Pi^{(0)}_{\mu\nu}  \mathcal{A}_{\mu\nu} = - 2 P_\perp^2 \, \mathcal{A}_0 \,,
&&
   \Pi^{(1)}_{\mu\nu}  \mathcal{A}_{\mu\nu} = \frac{(qq')}{\sqrt{-q^2}} P_\perp^2 \, \mathcal{A}_1\,,
&&
   \Pi^{(2)}_{\mu\nu}  \mathcal{A}_{\mu\nu} = P_\perp^4 \, \mathcal{A}_2 \,.
\end{align}
Neglecting ``genuine'' higher-twist contributions due to quark-gluon correlations, the amplitudes
$\mathcal A^{(k)}$ can be written as a convolution of the generalized parton distributions $H_q(x,\xi,t)$ and the coefficient functions
$T^{(k)}(u,Q^2,t)$
\begin{align}
\mathcal A^{(k)} &= T^{(k)}\otimes {H} ~ \overset{\text{def}}{=}~
\sum_{q} e_q^2 \int_{-1}^1 \frac{dx}{2\xi} T^{(k)}\left(\frac{\xi+x-i\epsilon}{2(\xi-i\epsilon)},Q^2,t\right)
\, H_q(x,\xi,t)\,.
\label{convolution}
\end{align}

Note that within our conventions $P_\perp^\mu$ \eqref{Pperp} is the only existing transverse four-vector so that it can
only be dotted onto itself. As a consequence,  the power expansion of the coefficient functions $T^{(k)}\left(z,Q^2,t\right)$
can conveniently be organized in terms of the two expansion  parameters
\begin{align}
                     \frac{t}{(qq')} \qquad\text{and}\qquad \frac{|\xi P_\perp|^2}{(qq')},
\label{expand22}
\end{align}
where $(qq')= -(Q^2+t)/2$. In this way the dependence of power corrections on the mass of the target  enters  only through the  dependence
on $t_{\mathrm min}$:
\begin{align}
|\xi P_\perp|^2 = \frac{1-\xi^2}{4} (t_{\mathrm min} -t) = - \xi^2 m^2 -\frac{1-\xi^2}4 t\,.
\label{expand23}
\end{align}
In what follows we will discuss the general structure of this expansion and derive explicit expressions for the first few terms.

%%%%%%%%%%%%%%%%%%%%%%%%%%%%%%%%%%%%%%%%%%%%%%%%%%%%%%%%%%%%%%%%%%%%%%%%%%%%%%%%%%%%%%%%%%%%%%%%%%%%%%%%%%%%%%%%%%%%%%%%%%%%%%%%%%%%%%%%%%%%%
\subsubsection{Generalized parton distributions}
%%%%%%%%%%%%%%%%%%%%%%%%%%%%%%%%%%%%%%%%%%%%%%%%%%%%%%%%%%%%%%%%%%%%%%%%%%%%%%%%%%%%%%%%%%%%%%%%%%%%%%%%%%%%%%%%%%%%%%%%%%%%%%%%%%%%%%%%%%%%%

The GPD $ H_q(x,\xi,t)$ is defined as a matrix element of the leading-twist light-ray operator
\begin{align}\label{Opprime}
\langle p^\prime|\mathcal {O}_q(z_1n,z_2n)|p\rangle = 2P_+
\int_{-1}^1 dx\, e^{-i P_+[z_1(\xi-x)+z_2(x+\xi)]}  H_q(x,\xi,t)\,,
\end{align}
where
\begin{align}\label{Oqop}
\mathcal O_q(z_1n,z_2n)=\frac12\Big(\bar q(z_1 n)\gamma_+ q(z_2 n)-\bar q(z_2 n)\gamma_+ q(z_1 n)\Big)\, .
\end{align}
Wilson lines between the quarks are implied,
and the ``plus'' projection is defined with respect to an arbitrary light-like vector
$P_+ = P_\mu n^\mu$, $n^2=0$. In what follows we omit the flavor index, $\mathcal O_q \to \mathcal O$.

On intermediate steps of the calculation, a particular version of the so-called double distribution (DD) representation
\cite{Radyushkin:1997ki,Radyushkin:1998bz} for this matrix element proves to be more convenient, see Ref.~\cite{Braun:2012bg}:
\begin{align}
\langle p'|\mathcal O(z_1n,z_2n)|p\rangle
=
\frac{2i}{z_{12}} \int_{-1}^1  d\beta \int^{1-|\beta|}_{-1+|\beta|} d\alpha\,
e^{-i(\ell_{z_1z_2}n)}\Phi(\beta,\alpha,t)\,,
\label{DD2}
\end{align}
where
\begin{align}
\ell_{z_1 z_2}^\mu=-z_1\Delta^\mu+(z_2-z_1)\Big[\beta P^\mu-\frac12(\alpha+1) \Delta^\mu\Big].
\end{align}
The DD $\Phi(\beta,\alpha,t)$ is symmetric under reflection $(\beta,\alpha)\mapsto (-\beta,-\alpha)$,
\begin{align}
\Phi(\beta,\alpha,t) & = \Phi(-\beta,-\alpha,t)\,
\end{align}
and can be represented as a total derivative~\cite{Teryaev:2001qm}
\begin{align}\label{singleDD}
\Phi(\beta,\alpha,t) & = \partial_\beta f(\beta,\alpha,t)+\partial_\alpha g(\beta,\alpha,t)\,.
\end{align}
As a consequence, the first moments of $\Phi(\beta,\alpha,t)$ vanish:
\begin{align}
\iint d\beta\, d\alpha \,\Phi(\beta,\alpha, t) ~=~
\iint d\beta\, d\alpha\, \alpha\, \Phi(\beta,\alpha, t)~=~
\iint d\beta\, d\alpha\,\beta\,  \Phi(\beta,\alpha, t) ~=~0\,,
\end{align}
where the integration regions are the same as in \eqref{DD2}.
This, in  turn,  guarantees that the r.h.s. of Eq.~\eqref{DD2} vanishes at $z_1\to z_2$.

The DD $\Phi(\beta,\alpha,t)$ and the GPD $H(x,\xi,t)$ are related as~\cite{Braun:2012bg}
\begin{align}\label{HPhi}
\partial_x H(x,\xi,t) &= \iint d\beta\, d\alpha\,\delta(x-\beta-\xi\alpha)\,\Phi(\beta,\alpha,t)\,.
\end{align}
Staying with the DD representation, our results for power corrections to helicity amplitudes
are given by a sum of terms of the following type
\begin{align}
I_{-1}(Y)& =\iint d\beta\, d\alpha\, \Phi(\beta,\alpha)\, Y(F) ,
\nonumber\\
I_k(Y) & =\iint d\beta\, d\alpha\, \Phi(\beta,\alpha)\,\beta\, (\beta\partial_F)^{k} Y(F), \qquad k=0,1,\ldots
\end{align}
where $Y(F)$ are certain functions of the variable
\begin{align}
F=\frac12\left(\frac\beta\xi+\alpha+1\right)\,, &&  F\, \delta(x-\beta-\xi\alpha) = \frac{x+\xi}{2\xi} \delta(x-\beta-\xi\alpha)\,.
\label{F}
\end{align}
These integrals can be rewritten in terms of the GPD $H(x,\xi,t)$:
\begin{align}
I_{-1}(Y) &= - \int_{-1}^1 \frac{dx}{2\xi} \, Y^\prime \left(\frac{x+\xi}{2\xi}\right)\, H(x,\xi)
~=~ - Y^\prime \otimes H\,, \qquad  Y^\prime(z) = \frac{d}{dz} Y(z)\,,
\notag\\
I_k(Y) &= - (-2 D_\xi)^{k+1}  \int_{-1}^1 \frac{dx}{2\xi} \, Y\left(\frac{x+\xi}{2\xi}\right)\, H(x,\xi)
~=~  - (-2 D_\xi)^{k+1} \big(Y \otimes H\big)\,,
\label{I(Y)}
\end{align}
where
\begin{align}
D_\xi \equiv \xi^2\partial_\xi \,.
\label{Dxi}
\end{align}
In this way all our results can be rewritten in the GPD representation which appears to be more suitable in applications.

Last but not least, the OPE for the product of two electromagnetic currents in Eq.\eqref{LROPE} is written in terms of the leading-twist
projection of the nonlocal quark-antiquark operator at a non-light-like separation $x^2\neq 0$ \eqref{LRoperator} implying
\begin{align}
\langle p'|\mathscr{O}(z_1,z_2)|p\rangle
=
\frac{2i}{z_{12}} \int_{-1}^1  d\beta \int^{1-|\beta|}_{-1+|\beta|} d\alpha\,
\big[e^{-i(\ell_{z_1z_2} x)}\big]_{lt}\Phi(\beta,\alpha,t)\,,
\label{DD3}
\end{align}
which involves the leading-twist projection of the exponential function~\cite{Balitsky:1987bk,Balitsky:1990ck}.
The definition of this function and some useful representations are presented in appendix~\ref{app:LTexp}.

The following scalar products are useful in the calculation:
\begin{align}
  q'\cdot \ell_{z_1z_2} &= - (q q') \big[ z_1  - z_{12} F\big],
\qquad
(\Delta\cdot  \ell_{z_1z_2}) = 
-\Delta^2 \left(z_1-z_{12}F + z_{12}\frac\beta\xi\right),
\notag\\
 \ell^2_{z_1z_2} &= -z^2_{12}\beta^2|P_\perp|^2 + \Delta^2 
\left(z_1-z_{12}F\right)\left(z_1-z_{12}F + z_{12}\frac\beta\xi\right),
\label{usefulscalar}
\end{align}
where $F$ is the variable defined in \eqref{F}.

%%%%%%%%%%%%%%%%%%%%%%%%%%%%%%%%%%%%%%%%%%%%%%%%%%%%%%%%%%%%%%%%%%%%%%%%%%%%%%%%%%%%%%%%%%%%%%%%%%%%%%%%%%%%%%%%%%%%%%%%%%%%%%%%%%%%%%%%%%%%%
\subsection{Helicity-flip amplitude $\mathcal{A}^{(2)}$} \label{sect:A2}
%%%%%%%%%%%%%%%%%%%%%%%%%%%%%%%%%%%%%%%%%%%%%%%%%%%%%%%%%%%%%%%%%%%%%%%%%%%%%%%%%%%%%%%%%%%%%%%%%%%%%%%%%%%%%%%%%%%%%%%%%%%%%%%%%%%%%%%%%%%%%

The remaining calculation is in principle straightforward but rather cumbersome because for $q'^2=0$
several individual contributions to the OPE \eqref{LROPE} suffer from infrared (IR) singularities.
When necessary, we use finite $|q'^2|\ll Q^2$ as the regulator.
We will find that all IR-divergent terms cancel in the sum so that the real photon limit can be taken at the end.
The helicity-flip amplitude  $\mathcal{A}^{(2)}$  proves to be the simplest. We choose this case for illustration.

Application of the projection operator \eqref{projectors} $\mathcal{A}_2 = \Pi^{(2)}_{\mu\nu}  \mathcal{A}^{\mu\nu} /P_\perp^4 $
eliminates all contributions $\sim g_{\mu\nu}, \Delta_\mu, \Delta_\nu$ and antisymmetric terms $\mu\leftrightarrow\nu$.
In addition, terms with $x_\mu\partial_\nu$ or  $x_\nu\partial_\mu$ can be rewritten using integration by parts, e.g.,
\begin{align}
  \int d^4x\, e^{-iqx} \frac{1}{[-x^2+i0]} x_\mu\partial_\nu f(x)
&\mapsto  - \int d^4x\, e^{-iqx} \frac{2 x_\mu x_\nu}{[-x^2+i0]^2} f(x)\,,
\end{align}
with the $g_{\mu\nu}$ and $\sim q_\nu$ contributions dropped thanks to the projector.
The general expression in \eqref{LROPE} thus simplifies to
\begin{align}
\mathcal A^{(2)} &= \frac{\Pi^{(2)}_{\mu\nu}}{\pi^2 P_\perp^4}
\int d^4x\, e^{+iq'x}
\Biggl\{ - 4 \frac{ x^\mu x^\nu}{x^6} \biggl[
\int_0^1 d u\,\langle p'|\mathscr O(\bar u x,0)|p\rangle+ \int_0^1 dv\,\langle p'|\mathscr O(x,vx)|p\rangle\biggl]
\notag\\
&\quad
 - \frac{x^\mu x^\nu}{x^4}(i\Delta \partial_x)
 \int_0^1\!\!du\!\int_0^{ \bar u}\!\!\! dv\,
\left( 2 \ln\bar\tau + 2 \ln\bar u + \frac32 + \frac12 \frac{v^2}{\bar v^2} + \frac{v}{\bar v} - \frac{2\tau}{\bar\tau}
\frac{1}{\bar v}
+ \frac12 \frac{v}{\bar v} \delta(u)
\right) \langle p'|\mathscr O(\bar u,v)|p\rangle
\notag\\
&\quad
- \frac{\Delta^2}{2}\frac{ x^\mu x^\nu}{x^4}
 \int_0^1\!\!du\!\int_0^{ \bar u}\!\!\! dv\,
\left( 2\frac{v}{\bar v}
 +\frac12
- \frac{2\tau}{\bar\tau}+ \frac12 \frac{v^2}{\bar v^2}
+ \frac12 \frac{v}{\bar v} \delta(u)
\right)
\langle p'|\mathscr O(\bar u,v)|p\rangle
\notag\\
&\quad
- \frac14
\frac{x^\mu x^\nu}{x^2}
\Big( (i\Delta\partial_x) +\frac12\Delta^2\Big)(i\Delta \partial_x)
\int_0^1\!\!du\!\int_0^{ \bar u}\!\!\! dv\,
\frac{v}{\bar v}\left( \frac2{\bar\tau}-1\right)
\langle p'|\mathscr O(\bar u,v) |p\rangle
\notag\\
&\quad
+ \frac12
\frac{x^\mu x^\nu}{x^2}
\Big((i\Delta\partial_x) +\frac14\Delta^2\Big)(i\Delta \partial_x)
\int_0^1\!\!du\!\int_0^{ \bar u}\!\!\! dv\,
\Big(\ln\bar\tau+\frac{2\tau}{\bar\tau}\Big)
\langle p'| \mathscr O (\bar u,v) |p\rangle
\Biggr\}.
\label{LROPE-A2}
\end{align}
In the general case the matrix elements \eqref{DD3} will involve
\begin{align}
\ell_{\bar u,v}^\mu = -\bar u \Delta^\mu - (\bar u-v) [\beta P^\mu -\tfrac12(\alpha+1) \Delta^\mu]\,,
\end{align}
and the projection will produce factors
\begin{align}
 \Pi^{(2)}_{\mu\nu} \ell^\mu_{\bar u,v} \ell^\nu_{\bar u,v} = (\bar u-v)^2 \beta^2 P_\perp^4\,.
\end{align}

%%%%%%%%%%%%%%%%%%%%%%%%%%%%%%%%%%%%%%%%%%%%%%%%%%%%%%%%%%%%%%%%%%%%%%%%%%%%%%%%%%%%%%%%%%%%%%%%%%%%%%%%%%%%%%%%%%%%%%%%%%%%%%%%%%%%%%%%%%%%%
\subsubsection{Leading-power contribution $1/Q^2$}
%%%%%%%%%%%%%%%%%%%%%%%%%%%%%%%%%%%%%%%%%%%%%%%%%%%%%%%%%%%%%%%%%%%%%%%%%%%%%%%%%%%%%%%%%%%%%%%%%%%%%%%%%%%%%%%%%%%%%%%%%%%%%%%%%%%%%%%%%%%%%

This contribution arises from the most singular terms $1/x^6$ in Eq.~\eqref{LROPE-A2} and is already $1/Q^2$ suppressed in
comparison to the helicity-conserving amplitude $\mathcal{A}^{(0)}$.
Using
\begin{align}
  \Pi^{(2)}_{\mu\nu}  i \int d^4x\, e^{iq'x} \frac{x^\mu x^\nu}{(-x^2+i0)^3} [e^{-i\ell x}]_{lt}
&=
\frac12 \pi^2 \Pi^{(2)}_{\ell\ell} \frac{1}{[-2(q'\ell)]},
\end{align}
one obtains
\begin{align}
\mathcal{A}_2^{1/Q^2} &=
 -2 \iint d\beta\,d\alpha\,  \Phi(\beta,\alpha) \, \beta^2
\biggl[
\int_0^1 d u\, \bar u \frac{1}{(q'\ell_{\bar u,0})}
+
\int_0^1 d v\, \bar v \frac{1}{(q'\ell_{1,v})}
\biggr]
\notag\\&=
  \frac{2}{(qq')} \iint d\beta\,d\alpha\,  \Phi(\beta,\alpha)\,  \beta^2
\biggl[
\int_0^1 d u\, \bar u \frac{1}{\bar u (1  - F)}
+
\int_0^1 d v\, \bar v \frac{1}{ 1  - \bar v  F}
\biggr]
\notag\\&=
\frac{2}{(qq')} \iint d\beta\,d\alpha\,  \Phi(\beta,\alpha)\,  \beta^2
 \partial_F \biggl[\frac{1-2F}{1- F} \ln F \biggr],
\end{align}
where we used that $F\mapsto 1-F$ under reflection $(\alpha,\beta) \mapsto (-\alpha,-\beta)$. Since
$\Phi(\beta,\alpha) =  \Phi(-\beta,-\alpha)$, only the symmetric terms in  $F\leftrightarrow 1-F$
have to be kept under the integral.

As the final step, using Eq.~\eqref{I(Y)} the result can be rewritten in terms of the GPD
\begin{align}
 \mathcal{A}_2^{1/Q^2} &= - \frac{8}{(qq')} D_\xi^2
\int_{-1}^1 \frac{dx}{2\xi} \,\frac{2x}{x-\xi}\ln \left(\frac{x+\xi}{2\xi}\right)\, H(x,\xi,t)
\notag\\&=
\frac{16}{Q^2+t}
\xi^3 \partial_\xi^2 \int_{-1}^{1}\!dx\, \frac{x}{x-\xi} \ln \left(\frac{x+\xi}{2\xi}\right)  H(x,\xi,t)\,.
\end{align}
This expression agrees with \cite[Eq.(120)]{Braun:2012bg} up to a factor two%
\footnote{The result in \cite[Eq.(85)]{Braun:2012bg} is correct, but a factor two was lost when going over to the GPD representation.}.
Note that the expansion naturally goes in powers of $(qq') = - (Q^2+t)/2$, hence we leave it in this form.

%%%%%%%%%%%%%%%%%%%%%%%%%%%%%%%%%%%%%%%%%%%%%%%%%%%%%%%%%%%%%%%%%%%%%%%%%%%%%%%%%%%%%%%%%%%%%%%%%%%%%%%%%%%%%%%%%%%%%%%%%%%%%%%%%%%%%%%%%%%%%
\subsubsection{Next-to-leading-power contribution $1/Q^4$}
%%%%%%%%%%%%%%%%%%%%%%%%%%%%%%%%%%%%%%%%%%%%%%%%%%%%%%%%%%%%%%%%%%%%%%%%%%%%%%%%%%%%%%%%%%%%%%%%%%%%%%%%%%%%%%%%%%%%%%%%%%%%%%%%%%%%%%%%%%%%%

The $1/Q^4$ contribution is due to the terms $1/x^4$ in the second and the third line in Eq.~\eqref{LROPE-A2}. This calculation is
equally simple. Consider the term $\sim (i\Delta \partial)$ first. To this end we need a Fourier integral
\begin{align}
\Pi^{(2)}_{\mu\nu} \,  i \int d^4x\, e^{iq'x} \frac{x^\mu x^\nu}{(-x^2+i0)^2} (i\Delta\cdot\partial_x) [e^{-i\ell x}]_{lt} &=
 8\pi^2 (q'\cdot\Delta)
\frac{\ell^2 \Pi^{(2)}_{\ell\ell} }{[-2(q'\ell)]^3}
- 4\pi^2 (\ell \cdot\Delta)
\frac{\Pi^{(2)}_{\ell\ell}  }{[-2(q'\ell)]^2}.
\end{align}
Changing variables
\begin{align}
 v=\bar u w\,, && \ell_{\bar u, v} = \bar u \ell_{1,w}\,, && \int_0^1\!\!du\!\int_0^{ \bar u}\!\!\! dv
    = \int_0^1\!\!dw \!\int_0^1\!\!\! du\,\bar u \,,
\label{substitute}
\end{align}
makes the $u$-integration trivial, so that we get
\begin{align}
 \mathcal{A}_2^{1/Q^4} & \ni
- \frac{1}{(qq')^2} \iint d\beta\,d\alpha\,  \Phi(\beta,\alpha)\,\beta^2
\int_0^1\!dw\,\frac{\bar w}{w}
\Big(1 +\frac{\ln\bar w}{w}\Big)
\biggl[
\frac{(\ell_{1,w}\cdot\Delta)}{(1-\bar w  F)^2} - \frac{\ell_{1,w}^2}{(1-\bar w F)^3}
\biggr].
\end{align}
A higher power of $(1-\bar w F)$ in the denominator of the second term is not a reason for worrying, because
\begin{align}
   \ell^2_{1,w} &=
\bar w^2 \beta^2 P_\perp^2 + t(1- \bar w F) 
(1-\bar w F + \bar w \beta/\xi)
\label{ell-1w}
\end{align}
so that in this term either  $1/(1-\bar w F)^3 \mapsto 1/(1-\bar w F)^2$, or an extra $\bar w^2$ factor arises,
which softens the behavior of the integral at $F\to 1$ equivalent to $x\to\xi$ seen from~\eqref{F}. As the result, this contribution does not have
a stronger singularity at $x\to \xi$ as compared to the leading $1/Q^2$ term.
One obtains after a little algebra,
\begin{align}
 \mathcal{A}_2^{1/{Q^4}}\! &\! \ni
\frac{-1}{(qq')^2} \iint d\beta\,d\alpha\,  \Phi(\beta,\alpha)\,
\beta \left(\frac{P_\perp^2}2(\beta\partial_F)^3 + \frac32 \frac{t}{\xi} (\beta\partial_F)^2 + 2 t (\beta\partial_F) \!\right)\!
\biggl[\frac{\Li_2(F)-\Li_2(1)}{1-F} + \ln F
\biggr].
\label{A2Q4-part1}
\end{align}
The term $\sim \Delta^2$ in the third line in Eq.~\eqref{LROPE-A2} is treated similarly, using
\begin{align}
  \Pi^{(2)}_{\mu\nu}  i \int d^4x\, e^{iq'x} \frac{x^\mu x^\nu}{(-x^2+i0)^2} [e^{-i\ell x}]_{lt}
=
- 4 \pi^2 \Pi^{(2)}_{\ell\ell} \frac{1}{[-2(q'\ell)]^2}.
\end{align}
One obtains
\begin{align}
\mathcal{A}_2^{1/Q^4} & \ni
- \frac{\Delta^2}{(qq')^2} \iint d\beta\,d\alpha\,  \Phi(\beta,\alpha) \beta^2
\partial_F
\biggl[\frac{\Li_2(F)-\Li_2(1)}{1-F} + \frac12 \frac{\ln F}{1-F}
\biggr].
\label{A2Q4-part2}
\end{align}
Adding \eqref{A2Q4-part1} and \eqref{A2Q4-part2},  and using the integrals in  \eqref{I(Y)}
we get
\begin{align}
\mathcal{A}_2^{1/Q^4} &=
\frac{8}{(qq')^2} \left(P_\perp^2 D_\xi^4  - \frac32 \frac{\Delta^2}{\xi} D_\xi^3
+ \frac32 \Delta^2 D_\xi^2 \right)
\int_{-1}^1 \frac{dx}{2\xi} \, \biggl\{ \frac{2\xi}{\xi-x}
\Big[\Li_2\left(\frac{x\!+\!\xi}{2\xi}\right)-\zeta_2\Big] + \ln \left(\frac{x\!+\!\xi}{2\xi}\right)  \biggr\}
\notag\\&\quad
\times H(x,\xi,t) +\frac{2 \Delta^2}{(qq')^2} D_\xi^2
\int_{-1}^1 \frac{dx}{2\xi} \, \biggl[ \frac{2x}{\xi-x} \ln \left(\frac{x\!+\!\xi}{2\xi}\right) \biggr] H(x,\xi,t)\,.
\end{align}

%%%%%%%%%%%%%%%%%%%%%%%%%%%%%%%%%%%%%%%%%%%%%%%%%%%%%%%%%%%%%%%%%%%%%%%%%%%%%%%%%%%%%%%%%%%%%%%%%%%%%%%%%%%%%%%%%%%%%%%%%%%%%%%%%%%%%%%%%%%%%
\subsubsection{Next-to-next-to-leading-power contribution $1/Q^6$ and beyond}
%%%%%%%%%%%%%%%%%%%%%%%%%%%%%%%%%%%%%%%%%%%%%%%%%%%%%%%%%%%%%%%%%%%%%%%%%%%%%%%%%%%%%%%%%%%%%%%%%%%%%%%%%%%%%%%%%%%%%%%%%%%%%%%%%%%%%%%%%%%%%

These contributions arise from terms $1/x^2$  in the last two lines in Eq.~\eqref{LROPE-A2} and are beyond our
target accuracy \eqref{intro:2}. In what follows we sketch their calculation, nevertheless, in order to reveal what appears to
be a general pattern of the complications that arise beyond the next-to-leading power.

Start with the terms $\sim \Delta^2(i\Delta \partial_x)$ that are somewhat simpler. The relevant Fourier integral reads
\begin{multline}
 i^2 \Delta^\xi \! \int\! d^4x\, e^{iq'x} \frac{x^\mu x^\nu}{(-x^2\!+\!i0)} \partial_\xi [e^{-i\ell x}]_{lt}  =
32 \pi^2 (\Delta\cdot\ell) \frac{\Pi^{(2)}_{\ell\ell}}{A^3}
- 96 \pi^2  (\Delta\cdot q')\Pi^{(2)}_{\ell\ell} \biggl\{ \frac{\ell^2}{A^4} \biggl(\ln \frac{A}{A+\ell^2} +  \ln \frac{A}{q'^2}\biggr)
\\
- \frac{\ell^2}{(A+\ell^2)^3}\biggl[ \frac{11}{6} \frac{1}{A} + \frac{17}{2} \frac{\ell^2}{A^2} + 10 \frac{\ell^4}{A^3} + \frac{11}{3}
\frac{\ell^6}{A^4}\biggr]\biggr\} + \mathcal{O}(q'^2), \label{bad-integral}
\end{multline}
where we use a shorthand notation $A = -2 (q'\ell)$. There are two major differences with what we had before. First, this integral is IR
divergent in the $q'^2\to 0$ limit so that we keep finite $q'^2$ in the last term in the first line as a regulator. Second, there is a
factor $1/(A+\ell^2)^3 = 1/(q'-\ell)^6$ and also a logarithmic term $\ln \frac{A}{A+\ell^2}$ that did not appear previously. Since $A =
\mathcal{O}(Q^2)$ and $\ell^2 = \mathcal{O}(\Delta^2, \xi^2P_\perp^2)$, the expansion $1/(A+\ell^2)^3 = 1/A^3 - 3 \ell^3/A^4 + \ldots$
generates a series of power corrections to all powers. This is in contrast to Fourier integrals that we have seen above in $\frac{x_\mu
x_\nu}{x^6}$ and $\frac{x_\mu x_\nu}{x^4}$ contributions, which only produce terms with a given power suppression $1/Q^2$ and $1/Q^4$,
respectively. Note that the IR divergent contribution $\sim  \ln \frac{A}{q'^2}$ multiplies $(\Delta q')/A^4 = \mathcal{O}(1/Q^6)$ and does
not appear in higher powers.

Using \eqref{bad-integral} and changing variables \eqref{substitute} it is possible to do the $u$-integration
explicitly. One finds that the IR-divergent terms $\sim \ln q'^2$ cancel thanks to
\begin{align}
& \int_0^{1}\!\! du \,
 \biggl\{ \left[\frac{v}{\bar v}\left( \frac2{\bar\tau}-1\right)\right]
-       \left[\ln\bar\tau+\frac{2\tau}{\bar\tau}\right]\biggr\}_{v=\bar u w}  = 0\,,
\end{align}
and one obtains
\begin{align}
\mathcal{A}_2^{1/Q^6} &\ni
- \frac{ 3  \Delta^2}{2 (qq')^3} \iint d\beta\,d\alpha\,  \Phi(\beta,\alpha)\,\beta^2
 \int_0^1\!\!dw\,\frac{\bar w \ell^2_{1,w}}{(1-\bar w F)^4}  \biggl\{
\left(\frac{1}{\bar w} + \frac{1}{w}\ln\bar w  \right)  + \mathcal{O}\biggl(\frac{\ell^2_{1,w}}{(qq')(1-\bar w F)}\biggr)\biggr\}.
\label{Q6:1}
\end{align}

The terms $\sim (i\Delta \partial_x)^2$ can be treated in the same manner. The relevant Fourier integral has similar structure
as in \eqref{bad-integral}, but is somewhat more cumbersome. The IR-divergent contributions $\sim \ln q'^2$
cancel also in this case, thanks to another identity
\begin{align}
& \int_0^{1}\!\! du \,\bar u
 \biggl\{ \left[\frac{v}{\bar v}\left( \frac2{\bar\tau}-1\right)\right]
    - 2\left[\ln\bar\tau+\frac{2\tau}{\bar\tau}\right]\biggr\}_{v=\bar u w}  = 0\,.
\end{align}
We obtain
\begin{align}
\mathcal{A}_2^{1/Q^6} &\ni
\frac{12}{(qq')^3} \iint d\beta\,d\alpha\,  \Phi(\beta,\alpha)\, \beta^2
 \int_0^1\!\!dw\,\frac{\bar w \ell^4_{1,w}}{(1-\bar w F)^5}
\biggl\{\left(\frac{1}{w^2} \ln\bar w +  \frac{1}{w} + \frac{2}{3\bar w} - \frac{1}{6}
\right)
\notag\\&\quad
+ \mathcal{O}\biggl(\frac{\ell^2_{1,w}}{(qq')(1-\bar w F)}\biggr)
\biggr\}.
\label{Q6:2}
\end{align}
One can show that each term in the expansion of the integrands in \eqref{Q6:1} and \eqref{Q6:2} in powers of $\ell^2/(qq')$
is $\mathcal{O}(w^1)$ at $w\to 0$, so that the
remaining integrals are convergent order by order in the power expansion.
Closed expressions for the integrands (to all powers) can be obtained, but are rather unwieldy%
\footnote{On can show that all further power corrections in these expressions (beyond $1/Q^6$)
 originate from large separations between the currents, of the order of
$|x^2| \sim 1/|q'^2|$. These corrections are finite, but it is not obvious whether they should or could be included in the
coefficient function of the GPD. This issue requires further study.}.

The remaining calculation is straightforward. As already mentioned above, terms with
increasing powers of $\ell^2_{1,w}/(1-\bar w F)$  do not give rise to stronger singularities at $x\to \xi$
because either the additional factors of $1/(1-\bar w F)$ is cancelled in the ratio, or a $\bar w^2$-factor appears which smoothens the behavior of the integral
at $F\to 1$, see Eq.~\eqref{ell-1w}. Thus collinear factorization is not endangered.

%%%%%%%%%%%%%%%%%%%%%%%%%%%%%%%%%%%%%%%%%%%%%%%%%%%%%%%%%%%%%%%%%%%%%%%%%%%%%%%%%%%%%%%%%%%%%%%%%%%%%%%%%%%%%%%%%%%%%%%%%%%%%%%%%%%%%%%%%%%%%
\subsection{Results}
%%%%%%%%%%%%%%%%%%%%%%%%%%%%%%%%%%%%%%%%%%%%%%%%%%%%%%%%%%%%%%%%%%%%%%%%%%%%%%%%%%%%%%%%%%%%%%%%%%%%%%%%%%%%%%%%%%%%%%%%%%%%%%%%%%%%%%%%%%%%%

The calculation of $\mathcal{A}^{(1)}$ proves to be of similar complexity, whereas $\mathcal{A}^{(0)}$ is more involved.
The general scheme of the calculation remains the same, but the cancellation of $\ln q'^2$ contributions in $1/Q^4$ corrections
in the latter case is more tricky as the expansion of  Fourier integrals at $q'^2\to 0$ sometimes leads to
logarithmic divergences at $u\to 1$ (in notation of the previous sections).
This divergent contribution has to be isolated and treated separately. In addition, power divergences $\sim 1/q'^2$ appear in the contributions
of the last two lines in Eq.~\eqref{LROPE}, but cancel in the sum.
The final expressions for all helicity amplitudes in the DVCS limit $q'^2=0$ are finite.

We obtain
\begin{subequations}
\label{Afinal}
\begin{align}
\label{A0final}
\mathcal{A}_0 &= 2 \left(1 + \frac{t}{4(qq')}\right) \,(T_0\otimes H)
\notag\\
&\quad
 -\frac{t}{(qq')} \, (T_1\otimes H) + \frac2{(qq')}\left( \frac{t}{\xi}  + 2 |P_\perp|^2 D_\xi\right) D_\xi \,(T_3\otimes H)
 \notag\\
&\quad
+\frac12 \frac{t^2}{(qq')^2} \,  (\widetilde T_1\otimes H)
  +\frac{4t}{(qq')^2}\left(\frac{t}{\xi} + 2 |P_\perp|^2 D_\xi \right) D_\xi (T_2\otimes H)
  \notag\\
  &\quad
  + \frac{2}{(qq')^2}\left(
\left(\frac{t}{\xi} + 2 |P_\perp|^2 D_\xi \right)^2 - 2 |P_\perp|^4 D_\xi^2
  \right) D_\xi^2\, (T_5\otimes H)\,,
\\[2mm]
\label{A1final}
\mathcal{A}_1 &=  - \frac{4 Q}{(qq')} D_\xi (T_1\otimes H)
\notag\\&\quad
+\frac{8 Q}{(q'q)^2}
\left(\frac{t}{\xi} + |P_\perp|^2 D_\xi \right)
D_\xi^2 (T_2\otimes H)
- \frac{4 Q t }{(q'q)^2} D_\xi (T_3\otimes H)\,,
\\[2mm]
\label{A2final}
\mathcal A_2
&=- \frac{8}{(qq')}\left(1 + \frac{t}{4(qq')}\right) D_\xi^2\, (\widetilde T_1\otimes H)
\notag\\
&\quad
+\frac{4}{(qq')^2}\left( 3 t - 3 \frac{t}{\xi} D_\xi -2|P_\perp|^2 D_\xi^2\right) D_\xi^2
\,(T_2\otimes H)\,.
\end{align}
\end{subequations}
Here $D_\xi = \xi^2\partial_\xi$ \eqref{Dxi} and the convolution $\otimes$ is defined in Eq.~\eqref{convolution}.
The same expressions are valid for a pseudoscalar target (pion) as well, up to an overall isospin factor, cf.~\cite{Braun:2012bg}.

The CFs that we encounter to NNLO power accuracy are
\begin{align}
T_0(u) &= \frac{1}{1-u}\,,
\nonumber\\
T_1(u) &= - \frac{1}{u} \ln(1-u)\,,
\nonumber\\
\widetilde T_1(u) &= \frac{1-2 u}{u}\ln (1-u)\,,
\nonumber\\
T_2(u) &=  \frac{\Li_2(u)-\Li_2(1)}{1-u} - \ln(1-u)\,,
\notag\\
T_3(u) &=  \frac{\Li_2(u)-\Li_2(1)}{1-u} - \frac{\ln(1-u)}{2u} = T_2(u) - \frac12 \widetilde T_1(u)\,,
\notag\\
T_5(u)&=\left(\frac{7}{2}-\frac{1}{2u}\right)\ln(1-u)-
\left(\frac{3}{1-u}-2\right)\Big(\Li_2(u)-\Li_2(1)\Big).
\label{Tfunctions}
\end{align}
They are analytic functions of $u$ with a cut from $1$ to $\infty$. 
Functions of higher transcendentality appear on intermediate steps of the calculation but cancel in the final expressions. 
The convolution integral~\eqref{convolution} contains the
CFs on the upper side of the cut: $T(u)\mapsto T(u+i\epsilon)$ for $x>\xi$. One finds
\begin{align}
\text{Im}T_0(u+i\epsilon) &= \pi \delta(1-u)\,,
\notag\\
\text{Im}T_1(u+i\epsilon) &= \frac\pi u\, \theta(u-1)\,,
\notag\\
\text{Im}\widetilde T_1(u+i\epsilon) &= \pi\frac{2u-1}u\, \theta(u-1)\,,
\notag\\
\text{Im}T_2(u+i\epsilon) &= \pi\left(\frac{\ln u}{1-u}+1\right)\, \theta(u-1)\,,
\notag\\
\text{Im}T_3(u+i\epsilon) &= \pi\left(\frac{\ln u}{1-u}+\frac1{2u}\right)\, \theta(u-1)\,,
\notag\\
\text{Im}T_5(u+i\epsilon) &= \pi\left[\Big(2-\frac 3{1-u}\Big)\ln u+\frac1{2u}-\frac72\right]\, \theta(u-1)\,.
\end{align}
In certain applications, e.g.~\cite{Lorce:2022tiq}, the expressions for the helicity amplitudes in the DD representation can be 
more useful, see Appendix~\ref{sect:DDrep}.

Note that factors of $|P_\perp|^2$ in \eqref{Afinal} always enter in combination with the second power of the derivative, $D^2_\xi$, 
which can be traced to the $\beta^2$ factor in the expression for  $\ell_{z_1z_2}^2 = -z_{12}^2\beta^2|P_\perp|^2 +\ldots $ \eqref{usefulscalar}.
 Since $D_\xi = \mathcal{O}(\xi)$, the expansion is organized in  powers of $\xi^2|P_\perp|^2/(qq') \propto \xi^2 m^2/Q^2 + \mathcal{O}(t/Q^2)$ as indicated in
Eqs.~\eqref{expand22}, \eqref{expand23}. For nuclear targets effectively $m \mapsto A m$ and $\xi \mapsto \xi/A$ so that the target mass
corrections are not enhanced as compared to the nucleon.

Note also that the convolutions $\widetilde T_1\otimes H$,  $T_2\otimes H$ and $T_5\otimes H$ contain contributions $\mathcal{O}(1/\xi)$ in the 
small-$\xi$ limit. These contributions, however, either cancel in the sum of all terms or are annihilated by applications of $D_\xi$, 
so that the power corrections have the same small-$\xi$ behavior as the leading terms.  

%%%%%%%%%%%%%%%%%%%%%%%%%%%%%%%%%%%%%%%%%%%%%%%%%%%%%%%%%%%%%%%%%%%%%%%%%%%%%%%%%%%%%%%%%%%%%%%%%%%%%%%%%%%%%%%%%%%%%%%%%%%%%%%%%%%%%%%%%%%%%
\section{Numerical estimates and discussion}
%%%%%%%%%%%%%%%%%%%%%%%%%%%%%%%%%%%%%%%%%%%%%%%%%%%%%%%%%%%%%%%%%%%%%%%%%%%%%%%%%%%%%%%%%%%%%%%%%%%%%%%%%%%%%%%%%%%%%%%%%%%%%%%%%%%%%%%%%%%%%

A detailed study of the numerical impact of kinematic power corrections goes beyond the tasks
of this paper. This calculation has to be done at the level of cross sections,
taking into account finite-$t$ and target mass effects to kinematic (e.g. phase space) factors
\cite{Belitsky:2010jw,Belitsky:2012ch} and including
the interference with the Bethe-Heitler process.
Besides, such a complete analysis is probably not warranted for the study case of a scalar target.

In this Section we follow Ref.~\cite{Braun:2012bg} and present numerical estimates for the kinematic power corrections
to the imaginary parts of the helicity amplitudes~\eqref{Afinal}.
To this end we use a model for the GPD $H(x,\xi, t)$ corresponding to the
$N=1$ ansatz from Ref.~\cite{Radyushkin:2011dh}.
It is based on the so-called single-DD description
which is defined  by the ``gauge-fixing'' condition
$$
\alpha f(\beta,\alpha,t)=\beta g(\beta,\alpha, t)\,,
$$
imposed on the DDs $f$ and $g$ in~\eqref{singleDD}, see Ref.~\cite{Radyushkin:2011dh} for more details. It is assumed that the DD $f$
takes a factorized form
\begin{align}
f(\beta,\alpha,t)=q(\beta,t) h(\beta,\alpha)\,.
\label{singleDDh}
\end{align}
Here $q(x,t=0)$ is a (quark) parton distribution which we take as
\begin{equation}
q(x,t)=\theta(x)\,x^{-a(t)}(1-x)^3 e^{B t}
\end{equation}
and
\begin{align}
h(\beta,\alpha)=\frac34\frac{(1-|\beta|)^2-\alpha^2}{(1-|\beta|)^3}.
\end{align}
The function $h(\beta,\alpha)$ satisfies the normalization condition $\int_{-1+|\beta|}^{1-|\beta|}d\alpha \,h(\beta,\alpha)=1$. Note that
we use $q(x)\sim (1-x)^3$ which is characteristic for the proton target, because this is the case that is most interesting phenomenologically.
For the pion one usually assumes $q(x)\sim (1-x)^{1\div 2}$.

In realistic models, see e.g. Ref.~\cite{Goloskokov:2006hr}, the $t$-dependence of the DD is often included through
the corresponding dependence of the valence quark Regge trajectory $a(t)=0.48+0.9\, \text{GeV}^{-2} t$. 
This dependence interferes with the finite-$t$ power corrections that are subject of this work, so that we do not take it into account in what follows 
and, for simplicity, set $a=1/2$. 
The overall multiplicative $e^{B t}$ factor cancels out in the ratios that will be considered.

The imaginary parts of the helicity amplitudes involve $H(x,\xi,t)$ in the region $x \geq \xi$ only.
In this region one obtains a compact expression~\cite{Radyushkin:2011dh}
\begin{align}\label{Hmodel}
H(x,\xi,t)\Big|_{x\geq \xi}=\frac{3x}{4\xi}\int_{\beta_1}^{\beta_2}\frac{d\beta}{\beta^{1+a(t)}}
\biggl[\bar\beta^2-\left(\frac{x-\beta}{\xi}\right)^2\biggr] e^{Bt},
\end{align}
where $\beta_1=(x-\xi)/(1-\xi)$ and  $\beta_2=(x+\xi)/(1+\xi)$.

\begin{figure}[t]
\begin{center}
\includegraphics[scale=0.50]{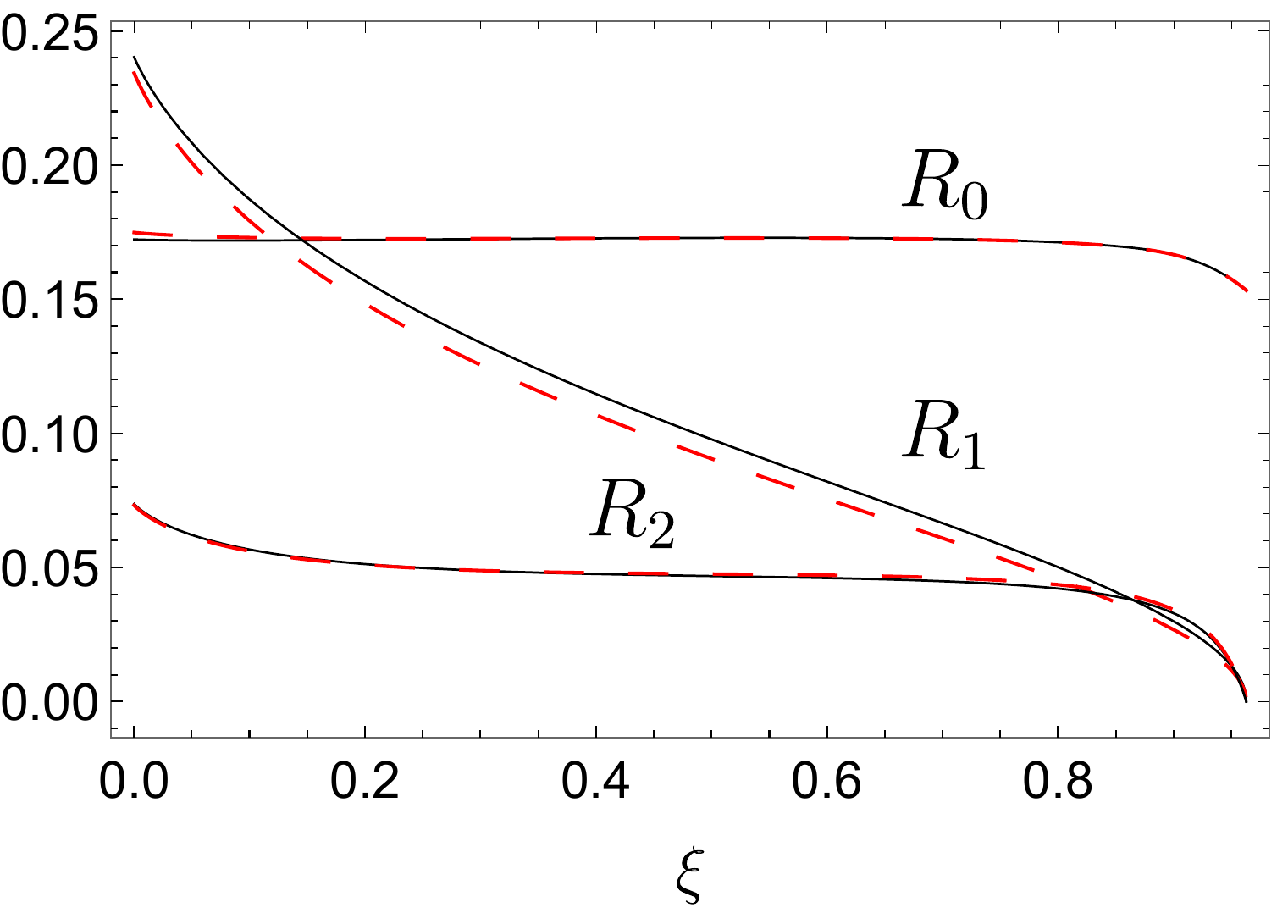}\hskip 7mm
\includegraphics[scale=0.50]{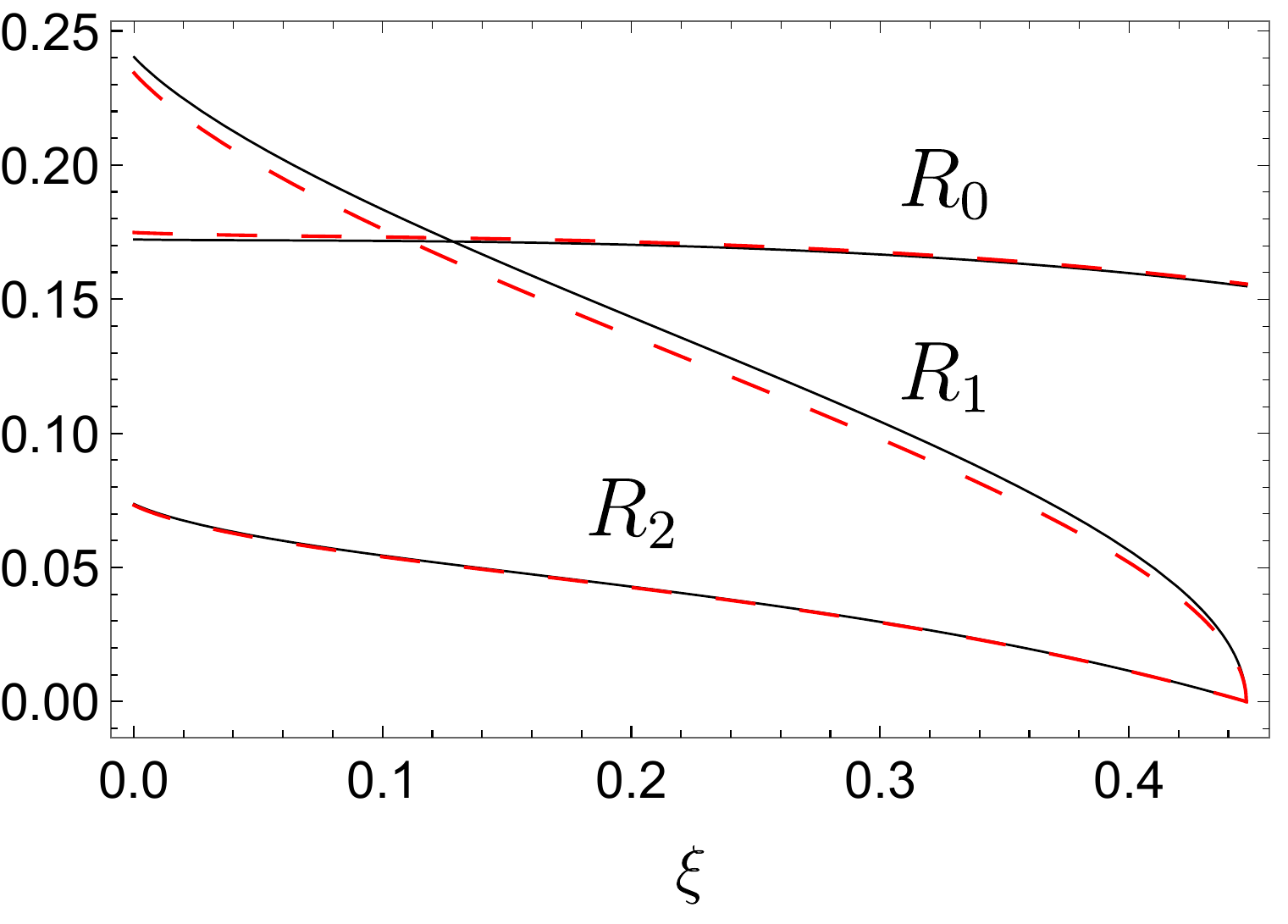}
\end{center}
\vspace{-5mm}
\caption{The ratios $R_k$ \eqref{R} of the imaginary parts of the helicity amplitudes taking into account 
leading kinematic power corrections (black solid curves) and the complete results \eqref{Afinal} to $1/(qq')^2$ accuracy (red dashed curves) 
as functions of the skewedness parameter for $Q^2=5$ GeV$^2$ and  $t=-1$ GeV$^2$.
The left panel: $m=0.14$ GeV; the right panel: $m=1$ GeV.
}
\label{fig:R}
\end{figure}

Kinematic power corrections modify the helicity-conserving amplitude $\mathcal{A}^{++} = \mathcal{A}_0$ 
and simultaneously give rise to helicity-flip contributions. In order to quantify both effects
we write the invariant functions $\mathcal{A}_k$  as  power series in $1/(qq')$ with ${\cal A}_k^{(p)} \sim 1/(qq^\prime)^p$
\begin{align}
    \mathcal{A}_0 &=  \mathcal{A}_0^{(0)} +  \mathcal{A}_0^{(1)} + \mathcal{A}_0^{(2)} +\ldots,
\notag\\
    \mathcal{A}_1 &=  \mathcal{A}_1^{(1)} +  \mathcal{A}_1^{(2)} +\ldots,
\notag\\
    \mathcal{A}_2 &=  \mathcal{A}_2^{(1)} +  \mathcal{A}_2^{(2)} +\ldots,
\end{align}
and plot in Fig.~\ref{fig:R} the ratios of the imaginary parts of the helicity amplitudes, see Eq.~\eqref{HelicityA}:
\begin{alignat}{3}
 R_0 &= \frac{\text{Im}  \mathcal{A}_0}{\text{Im}  \mathcal{A}^{(0)}_0} -1
&&\sim~ &&\frac{r_0^{(1)}}{(qq')}  + \frac{r_0^{(2)}}{(qq')^2} +\ldots\,,
\notag\\
 R_1 &= -\frac{|P_\perp|}{\sqrt{2}}\frac{\text{Im}  \mathcal{A}_1}{\text{Im}  \mathcal{A}^{(0)}_0}
&&\sim~ &&\frac{Q r_1^{(1)}}{(qq')}  + \frac{Q r_1^{(2)}}{(qq')^2} +\ldots\,,
\notag\\
 R_2 &= \frac12 |P_\perp|^2\frac{\text{Im}  \mathcal{A}_2}{\text{Im}  \mathcal{A}^{(0)}_0}
&&\sim~ && \frac{r_2^{(1)}}{(qq')}  + \frac{r_2^{(2)}}{(qq')^2} +\ldots\,,
\label{R}
\end{alignat}
normalized to the leading-twist contribution $\text{Im} \mathcal{A}_0^{(0)} = \pi H(\xi,\xi)$. 

The calculation is done for  $Q^2=5$ GeV$^2$,  $t=-1$ GeV$^2$
and two values of the the target mass: $m=0.14$ GeV (pion) with $m=1$ GeV (nucleon), see  Appendix~\ref{sect:numerics} for details. 
The results are presented on the left and the right panel in Fig.~\ref{fig:R}, 
respectively. 
The leading power contributions to the ratios $R_k(\xi)$ are shown by solid black curves and the complete results to the
$ 1/(qq')^2$ accuracy by red dashes. 

One sees that the contribution of subleading power corrections is small for all amplitudes. This is especially so for $R_0$ and $R_2$ where the 
difference between solid and dashed curves is within the line thickness. The smallness of the $1/(qq')^2$ corrections in these two cases is
due to strong cancellations between the several relevant contributions in the corresponding expressions in \eqref{Afinal}. This cancellation
apparently persists for a rather large class of the GPD models. Note, however, that the smallness of corrections only holds 
if the expansion is organized in powers of the scalar product $1/|(qq')| \sim 1/(Q^2+t)$ instead of $1/Q^2$. 
For the chosen values $Q^2=5$ GeV$^2$ and  $t=-1 $~GeV this is a 25\% effect.
                                                                               
The power correction to the leading, helicity-conserving amplitude $R_0$ depends very weakly on $\xi$ whereas
$R_1$ and $R_2$ vanish at the kinematically maximum allowed value of the skewedness parameter $ \xi = \xi_{\mathrm max}$ \eqref{tmin} 
owing to the $|P_\perp|$ factors in their definition.
The value of $\xi_{\mathrm max}$ depends strongly on the target mass, which explains the difference of the 
plots on the left (small mass) and right (large mass) panels. At small values of $\xi$ there is practically no difference, since,
as already mentioned earlier, the target mass corrections enter through the combination $\xi^2 m^2$ and become irrelevant at large energies.

%%%%%%%%%%%%%%%%%%%%%%%%%%%%%%%%%%%%%%%%%%%%%%%%%%%%%%%%%%%%%%%%%%%%%%%%%%%%%%%%%%%%%%%%%%%%%%%%%%%%%%%%%%%%%%%%%%%%%%%%%%%%%%%%%%%%%%%%%%%%%
\section{Conclusions}
%%%%%%%%%%%%%%%%%%%%%%%%%%%%%%%%%%%%%%%%%%%%%%%%%%%%%%%%%%%%%%%%%%%%%%%%%%%%%%%%%%%%%%%%%%%%%%%%%%%%%%%%%%%%%%%%%%%%%%%%%%%%%%%%%%%%%%%%%%%%%

Using the recent results \cite{Braun:2020zjm} on the contributions of descendants of the leading twist operators to the 
operator product expansion of two electromagnetic currents in conformal QCD, we have presented a calculation of finite-$t$ and target mass 
corrections to DVCS on scalar targets to the  next-to-leading power accuracy.
Our main result of phenomenological relevance is that the next-to-leading corrections are small if the expansion is reorganized in 
powers of $1/(Q^2+t)$ instead of $1/Q^2$. The calculation can be extended to higher powers. In particular we find that
IR divergences in kinematic corrections cancel to all powers to our present accuracy, in the leading order of perturbation theory.
We also argue that target mass corrections in the coherent DVCS from nuclei at large energies are small and do not invalidate the factorization theorem.

A generalization of these results to DVCS on spin-1/2 targets (nucleon) should be straightforward, but more tedious.
Also kinematic corrections to  double-DVCS (with two virtual photons) can be obtained. 
A more ambitious project would be to calculate kinematic corrections to the contribution of gluon GPD, that requires going over to 
next-to-leading order in the strong coupling.

%%%%%%%%%%%%%%%%%%%%%%%%%%%%%%%%%%%%%%%%%%%%%%%%%%%%%%%%%%%%%%%%%%%%%%%%%%%%%%%%%%%%%%%%%%%%%%%%%%%%%%%%%%%%%%%%%%%%%%%%%%%%%%%%%%%%%%%%%%%%%
\section*{Acknowledgments}
%%%%%%%%%%%%%%%%%%%%%%%%%%%%%%%%%%%%%%%%%%%%%%%%%%%%%%%%%%%%%%%%%%%%%%%%%%%%%%%%%%%%%%%%%%%%%%%%%%%%%%%%%%%%%%%%%%%%%%%%%%%%%%%%%%%%%%%%%%%%%
This work was supported in part by the Research Unit FOR2926 and 
the Collaborative Research Center TRR110/2 funded by 
the Deutsche Forschungsgemeinschaft (DFG, German Research Foundation) under grants 409651613 and 
196253076, respectively.

\appendix

%%%%%%%%%%%%%%%%%%%%%%%%%%%%%%%%%%%%%%%%%%%%%%%%%%%%%%%%%%%%%%%%%%%%%%%%%%%%%%%%%%%%%%%%%%%%%%%%%%%%%%%%%%%%%%%%%%%%%%%%%%%%%%%%%%%%%%%%%%%%%
\section{Derivation of Eqs.~\eqref{O1genfunction}, \eqref{O2genfunction}} \label{app:O2}
%%%%%%%%%%%%%%%%%%%%%%%%%%%%%%%%%%%%%%%%%%%%%%%%%%%%%%%%%%%%%%%%%%%%%%%%%%%%%%%%%%%%%%%%%%%%%%%%%%%%%%%%%%%%%%%%%%%%%%%%%%%%%%%%%%%%%%%%%%%%%

We start from the identities
\begin{subequations}
\label{O1O2}
\begin{align}\label{Oexp1}
\left[i P^{\alpha\dot\alpha},\partial_\alpha \bar\partial_{\dot\alpha} \partial_+^k \mathcal O^{(0)}_N\right ] &=
N^2 \partial_+^k \mathcal O^{(1)}_N
+\frac14 k(2N + k+1)\left[iP^{\alpha\dot\alpha},\left[iP_{\alpha\dot\alpha},\partial_+^{k-1} \mathcal O^{(0)}_N\right]\right],
\\
\label{Oexp2}
\left[i P^{\alpha\dot\alpha},\partial_\alpha \bar\partial_{\dot\alpha} \partial_+^k \mathcal O^{(1)}_N \right] &=
(N-1)^2 \partial_+^k \mathcal O^{(2)}_N
+\frac14 k(2N + k-1)\left[iP^{\alpha\dot\alpha},\left[iP_{\alpha\dot\alpha},\partial_+^{k-1} \mathcal O^{(1)}_N \right]\right],
\end{align}
\end{subequations}
where $P$ is the momentum operator and we use the two-component spinor notations as defined in Ref.~\cite{Braun:2011dg},
e.g. $P_{\alpha\dot\alpha}=P_\mu
(\sigma^\mu)_{\alpha\dot\alpha}$, $n_{\alpha\dot\alpha}=\lambda_\alpha \bar\lambda_{\dot\alpha}$,
$\partial_\alpha\equiv\frac{\partial}{\partial\lambda^\alpha}$, etc. In the matrix elements one can  replace $P\mapsto \Delta$ in
\eqref{O1O2}.

Multiplying both sides of \eqref{Oexp1} by $ \omega_{Nk} (S_+^{(1,1)})^k z_{12}^{N-1}$  and summing over $N$ and $k$ one obtains
\begin{align}\label{localO1expansion}
\sum_{N,k} \rho_N N^2 z_{12}^{N-1} \int_0^1 du(u\bar u)^N \mathcal O^{(1)}_N (n z_{21}^u) =
    \left(i\left(\Delta^{\alpha\dot\alpha}\partial_\alpha\partial_{\dot\alpha}\right) + \frac12\Delta^2 S_+^{(1,1)} \right)
        \mathscr O_{+}(z_1,z_2),
\end{align}
where $\rho_N$ and $\omega_{Nk}$ are defined in \eqref{rhoN} and \eqref{omegaNk}, respectively,
and $\mathscr O_{+}(z_1,z_2)$ is the light-ray operator for light-like separations,
\begin{align}
\mathscr O_{+}(z_1,z_2)=\mathscr O(z_1,z_2)|_{x\to n}, &&  \mathscr O(z_1,z_2)=\Pi(x,\lambda )\mathscr O_{+}(z_1,z_2)\,.
\end{align}
Here $\Pi(x,\lambda )$ is  the leading-twist projector, see Ref.~\cite[Eq.(5.26)]{Braun:2011dg}. Deriving \eqref{localO1expansion} we
take into account that $k(2N+k+1)\omega_{Nk}=\omega_{Nk-1}$ and
\begin{align}
\omega_{Nk} (S_+^{(1,1)})^k z_{12}^{N-1}=\rho_N \int_0^1 du (u\bar u)^N (z_{21}^u)^k.
\end{align}
Finally, applying the leading-twist projector to the both sides of \eqref{localO1expansion}
and taking into account that
\begin{align}\label{eq:Pidd}
\Pi(x,\lambda)\frac{\partial}{\partial \lambda^\alpha} \frac{\partial}{\partial\bar\lambda^{\dot\alpha}} \mathscr O_{+}(z_1,z_2)
 =\frac12 \partial_{\alpha\dot\alpha}\left(S_{0}^{(1,1)}-1\right)\Pi(x,\lambda )\mathscr O_{+}(z_1,z_2)
\end{align}
one ends up with the relation in Eq.~\eqref{O1genfunction}.

To derive \eqref{O2genfunction}
we start with \eqref{Oexp2}, multiply both sides by $ N^2 \omega_{Nk} (S_+^{(1,1)})^k z_{12}^{N-1}$, and sum over $N$ and $k$.
After some algebra one obtains
\begin{multline}
\sum_{N} \rho_N N^2 \int_0^1\! du\,(u\bar u)^N \left\{(N-1)^2 \mathcal O^{(2)}_N (n z_{21}^u) + \Delta^2 S_+^{(1,1)}
\int_0^1 \!dt\, t^{2N+1} \mathcal O^{(1)}_N (n t z_{21}^u)\right\} z_{12}^N
\\ =
\left(i\left(\Delta^{\alpha\dot\alpha}\partial_\alpha\partial_{\dot\alpha}\right) + \frac12\Delta^2 S_+^{(1,1)} \right) \sum_{N}
 \rho_N N^2 z_{12}^{N-1} \int_0^1\! du\,(u\bar u)^N \mathcal O^{(1)}_N (n z_{21}^u)\,.
\end{multline}
Applying the projector $\Pi(x,\lambda)$ to both sides one gets
\begin{multline}\label{A22e}
\sum_{N} \rho_N N^2 \int_0^1\! du\,(u\bar u)^N \left\{(N-1)^2 [\mathcal O^{(2)}_N (x z_{21}^u)]_{lt} + \Delta^2 S_+^{(1,1)}
\int_0^1 \!dt\, t^{2N+1}[ \mathcal O^{(1)}_N (n t z_{21}^u)]_{lt}\right\} z_{12}^N
\\ =
\left\{ \big(S_0^{(1,1)}-2\big) (i\Delta\partial_x) + \frac12 \Delta^2 S_+^{(1,1)}\right\} \Pi(x,\lambda) \sum_{N}
 \rho_N N^2 z_{12}^{N-1} \int_0^1\! du\,(u\bar u)^N \mathcal O^{(1)}_N (n z_{21}^u).
\end{multline}
Note the change from $S_0^{(1,1)}-1$ in~\eqref{eq:Pidd} to $S_0^{(1,1)}-2$ in the above equation. It happens because the spin of  the operator $\mathcal O^{(1)}_N$ is $N-1$, see definitions in~\eqref{O012}.
Finally, replacing the last sum in \eqref{A22e} by \eqref{O1genfunction} one arrives at Eq.~\eqref{O2genfunction}.

%%%%%%%%%%%%%%%%%%%%%%%%%%%%%%%%%%%%%%%%%%%%%%%%%%%%%%%%%%%%%%%%%%%%%%%%%%%%%%%%%%%%%%%%%%%%%%%%%%%%%%%%%%%%%%%%%%%%%%%%%%%%%%%%%%%%%%%%%%%%%
\section{Light-ray OPE: terms $\tfrac{x^\mu x^\nu}{x^4} [\mathscr{O}^{(1)}_{N}]_{lt}$} \label{app:example}
%%%%%%%%%%%%%%%%%%%%%%%%%%%%%%%%%%%%%%%%%%%%%%%%%%%%%%%%%%%%%%%%%%%%%%%%%%%%%%%%%%%%%%%%%%%%%%%%%%%%%%%%%%%%%%%%%%%%%%%%%%%%%%%%%%%%%%%%%%%%%

Here we illustrate our techniques on another example, the contributions $\sim \tfrac{x^\mu x^\nu}{x^4}\, [\mathscr{O}^{(1)}_{N}]_{lt}$.
There are two such terms: one is explicit in line seven (second to the last) of Eq.~\eqref{OPE} and another one arises from the second term
in the second line of Eq.~\eqref{OPE} when $\mathscr{O}^{(0)}_{N}$ is rewritten using \eqref{O0-lt} in terms of the leading-twist operators.
In the sum one obtains
\begin{align}
&- \frac{x^\mu x^\nu}{ x^4} \sum_{N>0,\text{even}}
\biggl[\frac{\rho_N N }{(N+1)}\biggr]
\int_0^1 \!du\, (u\bar u)^{N}
\biggl\{ \biggl[\frac{1}{N} - \frac{\bar u}{N+1} + B(N,\bar u)\biggr] +
\frac{\bar u-u}{N+1} \biggr\} [\mathscr{O}^{(1)}_{N}(ux)]_{lt}
\notag\\ &=
- \frac{x^\mu x^\nu}{ x^4} \sum_{N>0,\text{even}}
\biggl[\frac{\rho_N N }{(N+1)}\biggr]
\int_0^1 \!du\, (u\bar u)^{N}
\biggl\{\frac{1}{N} - \frac{u}{N+1} + B(N,\bar u)\biggr\} [\mathscr{O}^{(1)}_{N}(u x)]_{lt},
\end{align}
where
\begin{align}
  B(N,\bar u) = \bar u^{-N}\int_0^{\bar u} \frac{dv}{\bar v} v^{N+1}.
\end{align}
In this case it is convenient to write $\mathscr{O}^{(1)}_N(u x) $ as a formal Taylor series,
\begin{align}
  \mathscr{O}^{(1)}_N(u x)  \mapsto \sum_k \frac{d_{N}}{k!} u^k (i\Delta x)^k,
\label{app:local1}
\end{align}
which allows one to get rid on an unpleasant integral in $B(N,\bar u)$. 
%{\color{red}$d_N$ denotes the local operator ${\cal O}_N^{(1)}(y=0)$ as in~\eqref{O012} with $x_{12}=x$.} 
One obtains
\begin{align}
 &\int_0^1 \!du\, (u\bar u)^{N}
\biggl\{\frac{1}{N} - \frac{u}{N+1} + B(N,\bar u)\biggr\}  u^k
=
\frac{\Gamma(N+1)\Gamma(N+k+1)}{\Gamma(2N+k+2)}
\biggl\{\frac{1}{N(N+1)} + \frac{1}{N+k+1}\biggr\},
\end{align}
so that we get
\begin{align}
 - \frac{x^\mu x^\nu}{ x^4} \sum_{N,k} \frac{d_N}{k!} (i\Delta x)^k
\rho_N N^2 \frac{\Gamma(N+1)\Gamma(N+k+1)}{\Gamma(2N+k+2)}\biggl\{\frac{1}{N^2(N+1)^2} + \frac{1}{N(N+1)}\frac{1}{N+k+1}\biggr\}.
\end{align}
Now we can employ the operator identity \eqref{O1genfunction} where we set $z_1=z$, $z_2=0$. Using \eqref{app:local1} it becomes
 \begin{multline}
 \sum_{N,k} \frac{d_N}{k!} (i\Delta x)^k \rho_N N^2 z^{N+k-1} \frac{\Gamma(N+1)\Gamma(N+k+1)}{\Gamma(2N+k+2)}
\\
= \Big(z\partial_z+1\Big) (i\Delta\partial_x)\mathscr{O}(z,0)
+ \frac12 \Big( z^2 \partial_z +2 z\Big) \Delta^2 \mathscr{O}(z,0)\,.
\label{app:eq123}
\end{multline}
As explained in the text, extra factors $1/(N(N+1))^k$ can be emulated  by application of the invariant operator
$\mathcal{H}_+: T^{(1)}\otimes T^{(1)}\mapsto T^{(1)}\otimes T^{(1)}$:
\begin{align}
   [\mathcal{H}_+ f](z_1,z_2) &= \int_0^1\!d\alpha\!\int_0^{\bar\alpha}\!d\beta\, f(z_{12}^\alpha,z_{21}^\beta)\,,
\notag\\
   [\mathcal{H}^2_+ f](z_1,z_2) &= - \int_0^1\!d\alpha\!\int_0^{\bar\alpha}\!d\beta\, \ln(\bar\tau)\, f(z_{12}^\alpha,z_{21}^\beta)\,,\qquad
   \tau =\frac{\alpha\beta}{\bar\alpha\bar\beta}\, ,
\end{align}
%{\color{red} with ${\cal H}_+^2$ obviously emulates $1/(N^2(N+1)^2)$.}
The remaining factor $1/(N+k+1)$ can be eliminated by rescaling of the quark-antiquark separation.
    To see this, replace $z\to t z$ in Eq.~\eqref{app:eq123} and integrate
%{(note ${\mathscr O}(z_1,z_2)\to t {\mathscr O}(tz_1,tz_2)$ for $z\to tz$, see~\eqref{LRoperator})}:
\begin{eqnarray}
\lefteqn{\hspace*{-2cm}\int_0^1\!dt \, t \bigg\{\Big(z\partial_z+1\Big) (i\Delta\partial_x)\mathscr{O}(tz,0)
+ \frac12 t \Big( z^2 \partial_z +2 z\Big) \Delta^2 \mathscr{O}(tz,0)\biggr\} = }
\notag\\&=&
\int_0^1\!dt \, t \biggl\{ \sum_{N,k} \frac{d_N}{k!} (i\Delta x)^k \rho_N N^2 (tz)^{N+k-1} \frac{\Gamma(N+1)\Gamma(N+k+1)}{\Gamma(2N+k+2)}\biggr\}
\notag\\&=&
\sum_{N,k} \frac{d_N}{k!} (i\Delta x)^k \rho_N N^2 z ^{N+k-1} \frac{\Gamma(N+1)\Gamma(N+k+1)}{\Gamma(2N+k+2)} \frac{1}{N+k+1}\,.
\end{eqnarray}
Thus we get the contribution of the structure $\tfrac{x^\mu x^\nu}{x^4} [\mathscr{O}^{(1)}_{N}]_{lt}$,
\begin{align}\label{eq:example-xx}
\ldots &=   - \frac{x^\mu x^\nu}{ x^4}
 \bigg\{
 \Big(\mathcal{S}-1\Big) \mathcal{H}_+^2 (i\Delta\partial_x) \mathscr{O}(z,0)
+\frac12 \Delta^2  \mathcal{S} \mathcal{H}_+^2 \mathscr{O}(z,0)
\biggr\}\biggr|_{z=1}
\notag\\&\quad
- \frac{x^\mu x^\nu}{ x^4}
\int_0^1\!dt\, t\,
\bigg\{
  (i\Delta\partial_x) [ \Big(\mathcal{S}-1\Big) \mathcal{H}_+ \mathscr{O}](tz,0)
+\frac12 \Delta^2 t [\mathcal{S} \mathcal{H}_+ \mathscr{O}](tz,0)
\biggr\}\biggr|_{z=1},
\end{align}
where $\mathcal{S}:\,  T^{(1)}\otimes T^{(1)}\mapsto T^{(\frac32)}\otimes T^{(\frac12)}$ is the invariant operator
introduced in  Eq.~\eqref{1->3/2}. 
Following the argumentation in section~\ref{sect:Example}, we obtain
\begin{align}
  [\mathcal{S}\mathcal{H}_+ f](z_1,z_2) &=
 \int_0^{1}\!d\beta\, f(z_{1}, z_{21}^\beta)\,,
\notag\\
  [\mathcal{S}\mathcal{H}_+^2 f](z_1,z_2) &=
\int_0^1\!d\alpha\!\int_0^{\bar\alpha}\!d\beta\,\frac{\beta}{\bar\beta} f(z_{12}^\alpha, z_{21}^\beta)\,,
\label{app:eq125}
\end{align}
where from
\begin{align}
  \int_0^1\!dt\, t\, \Big[\mathcal{S}\,\mathcal{H}_+ f\Big](tz,0) &= \int_0^1\!d\alpha\, \int_0^{\bar\alpha}\!d\beta f(\bar\alpha z , \beta z)\,,
\notag\\
  \int_0^1\!dt\, t^2\,\Big[\mathcal{S} \, \mathcal{H}_+ f\Big](tz,0) &= \int_0^1\!d\alpha\,\bar\alpha\! \int_0^{\bar\alpha}\!d\beta f(\bar\alpha z ,
  \beta z)\,,
\notag\\
  \int_0^1\!dt\,t\, \Big[\mathcal{H}_+ f\Big](tz,0) &=
    - \int_0^1\!d\alpha\int_0^{\bar \alpha}\!d\beta\, \ln\bar\alpha\, f(\bar\alpha z , \beta z)\,.
\end{align}
Collecting everything, we end up with the desired expression
\begin{align}
- \frac{x^\mu x^\nu}{ x^4} \int_0^1\!d\alpha\int_0^{\bar \alpha}\!d\beta \biggl\{
\left(\frac{\beta}{\bar\beta} +  \ln\bar\tau  + 1 + \ln\bar\alpha \right)  (i\Delta\partial_x) \mathscr{O}(\bar\alpha, \beta)
+ \left(\frac{\beta}{\bar\beta} + \bar\alpha \right) \frac12 \Delta^2  \mathscr{O}(\bar\alpha, \beta)
\biggr\}.
\end{align}

%%%%%%%%%%%%%%%%%%%%%%%%%%%%%%%%%%%%%%%%%%%%%%%%%%%%%%%%%%%%%%%%%%%%%%%%%%%%%%%%%%%%%%%%%%%%%%%%%%%%%%%%%%%%%%%%%%%%%%%%%%%%%%%%%%%%%%%%%%%%%
\section{Leading-twist exponential function} \label{app:LTexp}
%%%%%%%%%%%%%%%%%%%%%%%%%%%%%%%%%%%%%%%%%%%%%%%%%%%%%%%%%%%%%%%%%%%%%%%%%%%%%%%%%%%%%%%%%%%%%%%%%%%%%%%%%%%%%%%%%%%%%%%%%%%%%%%%%%%%%%%%%%%%%

The leading-twist projection of the nonlocal quark-antiquark operator \eqref{LRoperator}
satisfies Laplace equation $\partial_x^2\mathscr{O}(z_1,z_2) = 0$ \cite{Balitsky:1987bk}, see section~\ref{sect:lt},
so that the expression on the r.h.s. of \eqref{DD3} must satisfy the same equation. Hence
\begin{align}
  \partial_x^2 \big[e^{-i\ell x}\big]_{lt} = \partial_\ell^2 \big[e^{-i\ell x}\big]_{lt} = 0\,
\end{align}
with the boundary condition that a usual exponential function is recovered if $x^2=0$ or $\ell^2=0$. The solution can be written as a power
series~\cite{Balitsky:1987bk}
\begin{align}
 [e^{-i\ell x}]_{lt} & = e^{-i\ell x} + \sum_{n=1}^\infty  \int_0^1\!dt\, (\tfrac14 x^2\ell^2)^n
\frac{t^n\, \bar t^{n-1}}{(n-1)! \,n!} e^{-i t \ell x},
\label{lt-exp1}
\end{align}
where in most applications only the first few terms are needed, cf.~\eqref{LTfunction1}.
Nevertheless, a closed expression summing all powers can be derived~\cite{Balitsky:1990ck}
\begin{align}
   [e^{-i(\ell x)}]_{lt} &= e^{-\frac{i}{2} (\ell x )}
\left[\cos \left(\frac12 r (\ell x) \right)
- \frac{i}{r}\sin \left(\frac12 r (\ell x) \right)\right],
\label{lt-exp2}
\end{align}
where
\begin{align}
 r = \sqrt{1 - \frac{\ell^2 x^2}{(x\cdot \ell)^2}}\,.
\end{align}
Note that the expansion of \eqref{lt-exp2} only involves even powers of $r$, so that there is no cut at $r=0$.

The Fourier transform of $[e^{-i(\ell x)}]_{lt}$ can be written in closed form as well,
\begin{align}
i \int \frac{d^4x}{\pi^2}\, \frac{ e^{iq'x}[e^{-i\ell x}]_{lt}}{[-x^2+i0]^p}
 &= \frac{\Gamma(2-p)}{2^{2p-3}\Gamma(p)}
\biggl[\left(1- \frac{(q'\ell)}{s_2}\right) (s_1-s_2)^{p-2}
%\notag\\&\quad
 +
      \left(1 + \frac{(q'\ell)}{s_2}\right) (s_1+s_2)^{p-2}
\biggr],
\end{align}
with
\begin{align}
  s_1 = (q'\ell) - q'^2\,, && s_2 = \sqrt{(q'\ell)^2 - \ell^2 q'^2}
\end{align}

%%%%%%%%%%%%%%%%%%%%%%%%%%%%%%%%%%%%%%%%%%%%%%%%%%%%%%%%%%%%%%%%%%%%%%%%%%%%%%%%%%%%%%%%%%%%%%%%%%%%%%%%%%%%%%%%%%%%%%%%%%%%%%%%%%%%%%%%%%%%%
\section{Helicity amplitudes in the DD representation}\label{sect:DDrep}
%%%%%%%%%%%%%%%%%%%%%%%%%%%%%%%%%%%%%%%%%%%%%%%%%%%%%%%%%%%%%%%%%%%%%%%%%%%%%%%%%%%%%%%%%%%%%%%%%%%%%%%%%%%%%%%%%%%%%%%%%%%%%%%%%%%%%%%%%%%%%
In this Appendix we present the expressions for the helicity amplitudes in the DD representation:
\begin{align}
\mathcal A_0 & =\! \iint\! d\alpha d\beta  \Phi(\beta,\alpha,t) \Biggl\{\!
\left(2+\frac{t}{2(qq')}\right) \ln(1-F) 
 + \frac{1}{(qq')}\left( t \Li_2(F) + \beta \left( \frac t\xi \!- {|P_\perp|^2}(\beta\partial_F) \right) T_3(F)  \right)
 \notag\\
&\quad
+
\frac{t^2}{(qq')^2}\left(\frac12 \Li_2(F)- (1- F) \ln(1- F)\right)
+\frac{2t}{(qq')^2}\beta \left( \frac t\xi- |P_\perp|^2 (\beta\partial_F) \right) T_2(F)
\notag\\
&\quad
+\frac{\beta}{(qq')^2}\left(-\frac{t^2}{2\xi^2} +t|P_\perp|^2\left(1+\frac1\xi (\beta\partial_F)\right)
-\frac{|P_\perp |^4}{4} (\beta\partial_F)^2\right)(\beta\partial_F) T_5(F)
\Biggr\},
\notag\\
\mathcal A_1 & =-\frac{2Q}{(qq')}\iint d\alpha d\beta \Phi(\beta,\alpha,t)\,\beta\Biggl\{ T_1(F)
+\frac{1}{(qq')}\biggl( t T_3(F) +\left( \frac t\xi-\frac{|P_\perp|^2}{2}(\beta\partial_F) \right)(\beta\partial_F) T_2(F)
\biggr)
\Biggr\},
\notag\\
\mathcal A_2 & =\frac 1{(qq')}\iint d\alpha d\beta \Phi(\beta,\alpha,t)\,\beta (\beta\partial_F)\Biggl\{
2\left(1+\frac{t}{4(qq')}\right) \widetilde T_1(F)
\notag\\
&\quad
-\frac 1 {(qq')}\left(
3t+\frac{3 t}{2\xi} (\beta\partial_F) -\frac{|P_\perp^2|}{2}(\beta\partial_F)^2\right)
T_2(F)
\Biggr\},
\end{align}
where $F = \tfrac12(\tfrac{\beta}{\xi}+\alpha +1)$ \eqref{F}
and the functions $T_i(F)$, $\widetilde T_1(F)$ are defined in~\eqref{Tfunctions}.

%%%%%%%%%%%%%%%%%%%%%%%%%%%%%%%%%%%%%%%%%%%%%%%%%%%%%%%%%%%%%%%%%%%%%%%%%%%%%%%%%%%%%%%%%%%%%%%%%%%%%%%%%%%%%%%%%%%%%%%%%%%%%%%%%%%%%%%%%%%%%
\section{Numerics}\label{sect:numerics}
%%%%%%%%%%%%%%%%%%%%%%%%%%%%%%%%%%%%%%%%%%%%%%%%%%%%%%%%%%%%%%%%%%%%%%%%%%%%%%%%%%%%%%%%%%%%%%%%%%%%%%%%%%%%%%%%%%%%%%%%%%%%%%%%%%%%%%%%%%%%%

The expressions \eqref{Afinal} for the amplitudes $\mathcal A_k$, $k=0,1,2$ contain derivatives with respect to $\xi$ up to the fourth
order. There are strong cancellations between the terms with different powers of $D_\xi$ in \eqref{Afinal}. This leads to a loss of
accuracy in numerical calculations. In order to avoid this problem it is preferably to bring the expressions for the amplitudes into the
form
\begin{align}\label{ImAF}
\text{Im}\,\mathcal A=\int dx \,F(x,\xi,t),
\end{align}
where the integrand $F$ receive contributions from terms with different powers of $D_\xi$. In order to do it we rescale $x\to x\xi$
in \eqref{convolution} and  write the convolution of the coefficient function and the GPD~\eqref{Hmodel} in the form:
\begin{align}\label{Jeta}
J(\eta)& =
\int_1^\eta dx V(x)\int_{\frac{x-1}{\eta-1}}^{\frac{x+1}{\eta+1}}
\frac{d\beta}{\beta^{1+a(t)}}
\biggl[\bar\beta^2-({x-\beta\eta})^2\biggr]\,,
\end{align}
where $\eta=1/\xi$ and
\begin{align}
V(x)=\frac38 x\, \text{Im}\, T\left(\frac{1+x}2\right)\,.
\end{align}
Since $D_\xi=-\partial_\eta$  we need to evaluate derivatives of $J(\eta)$ with respect to $\eta$. Taking the derivative of \eqref{Jeta}
one find that all boundary terms vanish and the final expression takes the form:
\begin{align}
\partial_\eta J(\eta)=2\int_1^\eta dx V(x)\int_{\frac{x-1}{\eta-1}}^{\frac{x+1}{\eta+1}}
\frac{d\beta}{\beta^{a(t)}}(x-\beta\eta) =2\int_1^\eta dx V(x)\left( x T_a(x,\eta)-\eta T_{a-1}(x,\eta) \right)\,,
\end{align}
where
\begin{align}
T_a(x,\eta) &=\int_{\frac{x-1}{\eta-1}}^{\frac{x+1}{\eta+1}}
\frac{d\beta}{\beta^{a(t)}}
= \frac{1}{1-a}\left(\left(\frac{x+1}{\eta+1}\right)^{1-a}-\left(\frac{x-1}{\eta-1}\right)^{1-a}\right)\,.
\end{align}
Similarly, one finds
\begin{align}
\partial_\eta^k J(\eta)=
2\int_1^\eta dx V(x)\partial_\eta^{k-1}\left( x T_a(x,\eta)-\eta T_{a-1}(x,\eta) \right) + \delta_{k4}
\frac{8V(\eta)}{(\eta^2-1)^2}
\,,
\end{align}
for $k=1,2,3,4$. It allows one to write the amplitudes in the form~\eqref{ImAF} and avoid the problem with  accuracy.
%%%%%%%%%%%%%%%%%%%%%%%%%%%%%%%%%%%%%%%%%%%%%%%%%%%%%%%%%%%%%%%%%%%%%%%%%%%%%%%%%%%%%%%%%%%%%%%%%%%%%%%%%%%%%%%%%%%%%%%%%%%%%%%%

\addcontentsline{toc}{section}{References}

\bibliography{references}

\bibliographystyle{JHEP}

%%%%%%%%%%%%%%%%%%%%%%%%%%%%%%%%%%%%%%%%%%%%%%%%%%%%%%%%%%%%%%%%%%%%%%%%%%%%%%%%%%%%%%%%%%%%%%%%%%%%%%%%%%%%%%%%%%%%%%%%%%%%%%%%

\end{document}